\documentclass[fleqn,usenatbib]{mnras}

\usepackage{newtxtext,newtxmath}

\usepackage[T1]{fontenc}

\DeclareRobustCommand{\VAN}[3]{#2}
\let\VANthebibliography\thebibliography
\def\thebibliography{\DeclareRobustCommand{\VAN}[3]{##3}\VANthebibliography}

\usepackage{graphicx}	
\usepackage{amsmath}	
\usepackage{subcaption}
\usepackage{array}
\usepackage{booktabs}
\usepackage[version=4]{mhchem}
\usepackage[symbol]{footmisc}
\usepackage{xcolor}

\title[{A Statistical and Multiwavelength Analysis of  IC 1590}]{A Statistical and Multiwavelength Photometric Analysis of a Young Embedded Open Star Cluster: IC 1590 }

\author[A. H. Sheikh]{
A. H. Sheikh$^{1}$\thanks{E-mail:\href{asheikh@gauhati.ac.in}{asheikh@gauhati.ac.in}}
Biman J. Medhi,$^{1}$\thanks{E-mail:\href{biman@gauhati.ac.in}{biman@gauhati.ac.in}}
\\
$^{1}$Department of Physics, Gauhati University, Guwahati 781014, Assam, India\\
}

\date{Accepted 2024 February 09. Received 2024 February 09; in original form 2023 December 11}

\pubyear{2023}

\begin{document}
\label{firstpage}
\pagerange{\pageref{firstpage}--\pageref{lastpage}}
\maketitle

\begin{abstract}
We present a statistical and multiwavelength photometric studies of young open cluster IC 1590. We identified 91 cluster members using $Gaia$ DR3 astrometry data using ensemble-based unsupervised machine learning techniques. From $Gaia$ EDR3 data, we estimate the best-fitted parameters for IC 1590 using the Automated Stellar Cluster Analysis package (ASteCA) yielding the distance $d$ $\sim$ 2.87 ± 0.02 Kpc, age $\sim$ 3.54 ± 0.05 Myr, metallicity $z$ $\sim$ 0.0212 ± 0.003, binarity value of $\sim$ 0.558, and extinction $A_v$ $\sim$ 1.252 ± 0.4 mag for an $R_v$ value of $\sim$ 3.322 ± 0.23. We estimate the initial mass function slope of the cluster to be $\alpha$ = 1.081 ± 0.112 for single stars and $\alpha$ = 1.490 ± 0.051 for a binary fraction of $\sim$ 0.558  in the mass range 1 M$_{\odot}$ $\leq$ m(M$_{\odot}$) $\leq$ 100 M$_{\odot}$. The $G$-band luminosity function slope is estimated to be $\sim$ 0.33 ± 0.09. We use $(J-H)$ versus $(H-K_s)$ color-color diagram to identify young stellar objects (YSOs). We found that all the identified YSOs  have ages $\leq$ 2 Myr and masses $\sim$ 0.35 - 5.5 M$_{\odot}$. We also fit the radial surface density profile. Using the galpy we performed orbit analysis of the cluster. The extinction map for the cluster region has been generated using the PNICER technique and it is almost similar to the dust structure obtained from the 500 $\mu$$m$ dust continuum emissions map of $Herschel$ SPIRE. We finally at the end discussed the star formation scenario in the cluster region.

\end{abstract}



\section{Introduction}

Studying young open clusters is essential for understanding the
processes involved in star formation. Open clusters are important stellar systems that provide valuable insights into the formation and evolution of stars. Open clusters are the group of stars that are gravitationally bounded and located relatively close to each other in space. They formed out of the same molecular cloud, so the members of an open cluster formed under similar physical conditions and at nearly the same timescale. In particular, the members of an open cluster have roughly the same distance, age, metallicity, luminosity function, mass function (IMF), etc. as they have the same origin. These fundamental parameters are the basic information that is needed to understand the formation and dynamical evolution of star clusters \citep{1986MNRAS.220..383S,2003ARA&A..41...57L}. Identifying cluster members from field stars (either background or foreground) that are stars not associated with the cluster, is essential in determining reliable results. This is especially important since the majority of young open clusters are situated within the Galactic plane \citep{2021AJ....162..140K}. So, to identify cluster members from the field stars, astrometric data like proper motions and parallaxes are very useful. By observing and analyzing open clusters, we can investigate the distribution of stellar masses, the effects of the cluster environment on star formation, and the dynamics of young stellar systems. Molecular clouds in the galactic disk fuel star formation by providing raw materials, such as molecular hydrogen, leading to gravitational collapse and the birth of stars, with their properties revealing insights into the processes shaping galactic stellar populations.\\
The young open cluster IC 1590 is embedded in an HII region and located
in the nebulosity of NGC 281 (Sharpless 184, PacMan Nebula, RA ($\alpha$) = $\text{00}^{h}\,\text{49}^{m}.9$, DEC ($\delta$) = $\text{+}56^{\circ} 21'$ and galactic longitude ($l$) = $123.^{\circ} 1$, galactic latitude ($b$) = -$06.^{\circ} 2$) \citep{1997AJ....113.2116G}. This NGC 281 complex is situated 300 pc above the Galactic plane, and it appears to be part of a 270 pc diameter ring of atomic and molecular clouds that is expanding at 22 km/s \citep{2003RMxAC..15..151M}. The most prominent feature of IC 1590 is the Trapezium-like system HD 5005 composed of four O-type stars HD 5005A (O4V(fc)), HD 5005B (O9.7II-III), HD 5005C (O8.5V(n)) and HD 5005D (O9.5V) \citep{2011ApJS..193...24S}. These massive stars are the primary ionizing sources of this HII region \citep{2021AJ....162..140K}. The cluster is known to be very young (3.5 - 4.4 Myr) and at a distance 2.0 - 3.5 Kpc from the Sun \citep{1997AJ....113.2116G,1994A&A...288..282H}. IC 1590 is found to be an interesting young cluster containing three emission-line stars belonging to, respectively, Herbig Ae, Herbig Be, and
classical Be \citep{2010BASI...38...35M}. Young open clusters are ideal for studying the evolution of disks of pre-main-sequence (PMS) stars, as they contain a significant number of PMS stars with circumstellar disks. In general, PMS stars are young low-mass sources that remain deeply embedded in the parent molecular cloud \citep{2021MNRAS.504..101L}.\\
Star formation can be triggered by the presence of massive stars in
young clusters. If massive stars with a spectral type earlier than B3 are formed within a molecular cloud, they emit intense ultraviolet (UV) radiation. This radiation is so powerful that it can ionize the hydrogen gas in their vicinity, causing it to become a hot and glowing region known as an HII region. Triggered star formation is a very complex process where the creation of massive stars affects their surroundings. These massive stars release energy that can heat and compress the leftover gas and dust from which they formed \citep{2012PASJ...64..107S,2021AJ....162..140K}. These massive stars emit strong stellar winds during their lifetime and eventually, when these stars explode in a supernova event, can push away the material surrounding them. In space, this can create a huge void-like structure called superbubbles or a galactic chimney \citep{2021AJ....162..140K}. As the size of these superbubbles increases, it can accumulate material, comprising dust and gas, in the peripheral regions, and may
induce subsequent star formation \citep{2008hsf1.book..459B}. So, there are two modes of triggered star formation happening here: a supernova initiates the cloud complex and the initial O-type stars, and then low mass star formation is subsequently triggered by the compression of molecular cores due to photoevaporation \citep{2003RMxAC..15..151M}. The NGC 281 region is a valuable laboratory for studying the triggered star formation through the interaction of massive stars with their surrounding material \citep{2012PASJ...64..107S}. This region contains various X-ray and IRAS sources, as well as an H$_2$O maser source, indicating ongoing star formation. So, this star formation may have been triggered by the interaction between the molecular cloud and the HII region of NGC 281 \citep{1978ApJ...219..467E,1997AJ....114.1106M}.

This work is organized as follows: In Section \ref{sec:data} we describe the data sources utilized for the IC 1590 open cluster, including $Gaia$ DR2, EDR3, and DR3 databases \citep{2018A&A...616A...1G,2021A&A...649A...1G,2023A&A...674A...1G}, the 2MASS Point Source Catalog (PSC) \citep{2003yCat.2246....0C},  the 3rd MSFRs Omnibus X-ray Catalog (MOXC3) \citep{2020yCat..22440028T}, and the SOS. VII. UBVI photometry of IC 1590 \citep{2021yCat..51620140K}. In Section \ref{sec:mem} we determine the cluster membership of IC 1590 using ensemble-based unsupervised machine learning techniques \citep{10.1093/mnras/stac2116}. Section \ref{sec:multi} contains the multiwavelength photometric analysis of IC 1590 to study the cluster’s stellar population and cluster parameters such as age, distance, metallicity, extinction, luminosity function, mass function (IMF), etc. Then we discuss the visual extinction map generated using the \href{http://smeingast.github.io/PNICER/ }{PNICER} technique \citep{2017A&A...601A.137M} from NIR $JHK_s$ data and compare the extinction structure to the dust structure obtained from the 500 $\mu$$m$ dust continuum emissions map of $Herschel$ SPIRE. Estimation of the distance of each cluster member using $Gaia$ DR2 parallaxes is discussed in Section \ref{sec:dis}. We describe the determination of the structural parameters of IC 1590 by employing radial surface number density fitted using a \citet{1962AJ.....67..471K} profile in Section \ref{sec:king}. In Section \ref{sec:orb} we estimate the birth radius and describe the orbit analysis of IC 1590 using \href{https://github.com/jobovy/galpy}{galpy}, a software package for galactic dynamics. Results thus obtained from this study alongside comparisons with previous studies have been presented in Section \ref{sec:res}. We finally discussed the summary and conclusions of this study in Section \ref{sec:sum}.
\section{Archival Datasets}
\label{sec:data}
\subsection{Gaia Data}
We used Gaia Data Release 3 ($Gaia$ DR3) \citep{2023A&A...674A...1G} for the cluster membership analysis in our study. The $Gaia$ DR3 is a significant astrometric and photometric catalog that provides precise astrometric solutions for over 1.468 billion sources \citep{2021A&A...649A...2L}. This release includes positions ($\alpha, \delta$), parallaxes, and proper motions ($\mu$$_{\alpha}$cos $\delta$, $\mu$$_{\delta}$) for these sources, with additional radial velocity measurements available for a subset of stars. The catalog offers photometric data in three bands: $G$, $G_{BP}$, and $G_{RP}$. It covers a wide range of magnitudes, from $G$ = 3 mag (bright sources) to a limiting magnitude of $G$ = 21 mag (faint sources). Systematic errors in astrometry have been significantly reduced in $Gaia$ DR3. Parallax uncertainties range from 0.02 to 0.03 mas for sources with $G$ < 15 mag and reach up to 1.3 mas at $G$ = 21 mag. Uncertainties in proper motion range from 0.02 to 0.03 mas yr$^{-1}$, but can reach 1.4 mas yr$^{-1}$ at $G$ = 21.

We have also used Gaia Early Data Release 3 ($Gaia$ EDR3) \citep{2021A&A...649A...1G} for the estimation of best-fitted fundamental parameters using  \href{https://github.com/asteca}{ASteCA} \citep{2015A&A...576A...6P}.  It has been known that the $Gaia$ EDR3 parallaxes have a global zero-point offset of $\sim$ 0.017 mas \citep{2021A&A...649A...2L}.

We used Gaia Data Release 2 ($Gaia$ DR2) \citep{2018A&A...616A...1G} for the distance estimation from parallax. The $Gaia$ DR2 provides a five-parameter astrometric solution: positions ($\alpha, \delta$), parallaxes, and proper motions for over 1.3 billion sources. The dataset extends its coverage up to a limiting magnitude of $G$ = 21, with a bright limit at $G$ = 3. Parallax uncertainties exhibit a range of up to 0.04 mas for sources with $G$ < 15, approximately 0.1 mas for sources with $G$ = 17, and, at the fainter end, the uncertainty reaches around 0.7 mas for sources at $G$ = 20.
\subsection{2MASS NIR JHK$_s$ Data}
We used near-infrared (NIR) $JHK_s$ point source data for the open cluster IC 1590 to identify YSOs in the cluster. NIR data has been obtained within a search radius of 9 arcmin from the Two Micron All Sky Survey (2MASS) Point Source Catalogue \citep{2003yCat.2246....0C} that contains astrometry and photometry in the three survey bandpasses: [$J$ (1.235 $\mu m$), $H$ (1.662 $\mu m$), and $K_s$ (2.159 $\mu m$)] for 470,992,970 sources.  The data is reported as 99 percent complete up to 16, 15, and 14.7 mag in the $J$, $H$, and $K_s$  photometric bands, respectively. We used a radius of 9 arcmin essentially the same as \citet{2012PASJ...64..107S} who estimated the cluster radius from the optical data as well as from the NIR 2MASS data as $R_{cl}$ $\sim$ 8 arcmin. To ensure accurate photometry, we only used data with ph-qual = $AAA$, indicating $S/N$ $\geq$ 10 and photometric uncertainty $\leq$ 0.10 \citep{2007MNRAS.379.1237M}.
\subsection{MOXC3 Chandra X-ray Data}
We used the third installment of the Massive Star-forming Regions (MSFRs) Omnibus X-ray Catalog (MOXC3) \citep{2020yCat..22440028T} to identify X-ray sources within the search radius of 9 arcmin, which is a compilation of X-ray point sources detected in 50 archival $Chandra$ Advanced CCD Imaging Spectrometer observations of 14 Galactic MSFRs and surrounding fields. It includes a rough limiting luminosity ($L_{tc}$) and the corresponding limiting mass ($M_{50\%}$) where the brighter half of the pre-main-sequence (pre-MS) X-ray population is sampled, based on results from the $Chandra$ Orion Ultradeep Project \citep{Preibisch_2005}.
\subsection{SOS. VII. UBVI Photometry Data}
We used the SOS. VII. $UBVI$ Photometry Catalogue of Open Cluster IC 1590 \citep{2021yCat..51620140K} to identify H$_{\alpha}$ sources within the search radius of 9 arcmin. The observations were made on October 31, 2011, using the Kuiper 61'' telescope (f/13.5) and Mont4k CCD of the Steward Observatory at Mt. Bigelow in Arizona for UBVI and H$_\alpha$ observations. The pixel scale is 0$''$.42 pixel$^{-1}$ in a 3$\times$3 binning
mode and the FoV is about 9$'$.7$\times$9$'$.7 \citep{2021AJ....162..140K}.
\section{Open Cluster Membership Analysis}
\label{sec:mem}
We have applied an improved method for the estimation of membership probability of open cluster IC 1590 using ensemble-based unsupervised machine learning techniques, which is very efficient and powerful in segregating the cluster members from the field stars \citep{10.1093/mnras/stac2116}. This technique involves the following two steps:
\par (i) The first step involves the selection of an appropriate range of cluster astrometric parameters. This selection is carried out using parallax ($\pi$), proper motion in right ascension ($\mu_{\alpha}$$\cos \delta$), and proper motion in declination  ($\mu_{\delta}$) data downloaded from $Gaia$ Archive\footnote{\label{fn:ex}\url{https://gea.esac.esa.int/archive/}} for IC 1590, within a smaller search of 3 arcmin radius by employing $k$-nearest neighbor (kNN) algorithm.
\par (ii) In the second step, a Gaussian Mixture Model (GMM) with two components is employed on the derived Gaussian distribution of Mahalanobis distance (MD) for stars. This utilizes the specified range of three astrometric parameters ($\pi$, $\mu_{\alpha}$$cos \delta$, $\mu_{\delta}$) acquired in the previous step. However, data from the $Gaia$ Archive$^{\ref{fn:ex}}$ is now downloaded within a larger search of 9 arcmin radius. The MD is computed based on the mean of each of the parameters in three dimensions. This approach effectively reduces the input dimensionality of the GMM from three-dimensional (3D) parameter space to one-dimensional (1D) parameter space, facilitating the determination of cluster membership.\\
The outcome of the 1D Gaussian distribution is determined using a two-component Gaussian Mixture Model (GMM) with the Expectation-Maximization (EM) algorithm \citep{10.1093/mnras/stac2116}. This approach is employed to determine the membership probabilities of the open cluster.
\subsection{kNN Technique for Outlier Removal}
\label{sec:knn}
The kNN algorithm \citep{1053964} is used to remove the most probable outliers among a set of stars, specifically those that are considered to be field stars, based on a cut-off (threshold) value of the average nearest neighbor distance ($\overline{d}_{\text{NN}}$) of the stars in the sample. $\overline{d}_{\text{NN}}$ is calculated by averaging the Euclidean distances between each star and its closest neighbors within a three-dimensional parameter space of ($\pi$, $\mu{\alpha}$$\cos \delta$, $\mu_{\delta}$). This step, called outlier removal, helps identify data points that are significantly distant from their neighbors, indicating potential outliers \citep{10.1093/mnras/stac2116}. The average nearest neighbor distance, $\overline{d}_{\text{NN}}$ is expressed as,
\begin{equation}
\overline{d}_\text{NN} = \sum_{k} \frac{d(x,k)}{{NN}_k}
\label{eq:knn1}
\end{equation}
where  $d(x, k)$ represents the Euclidean distance of a data point $x$ from its $k^{th}$ nearest neighbors, with $NN_k$ denoting the total nearest neighbors, where $k \in NN_k$. For cluster stars, $\overline{d}{\text{NN}}$ is expected to be smaller compared to field stars. A star is identified as an outlier if $\overline{d}{\text{NN}} > t$, where $t$ is the threshold (cut-off) value \citep{10.1093/mnras/stac2116}. The Euclidean distance $D_E(x, y)$ between two data points $x$ and $y$ in an n-dimensional space can be expressed as \citep{10.1093/mnras/stac2116},
\begin{equation}
D_{E}(x,y) = \sqrt{\sum_{i=1}^{n} {(x_{i}-y_{i}})^2}
\label{eq:knn2}
\end{equation}
where $n$ denotes the number of dimensions in the data, while $x_{i}$ and $y_{i}$ represent the $i^{th}$ components of points $x$ and $y$ respectively.

\begin{figure*}
\centering
\begin{subfigure}[b]{0.32\textwidth}
\centering
\includegraphics[width=\textwidth]{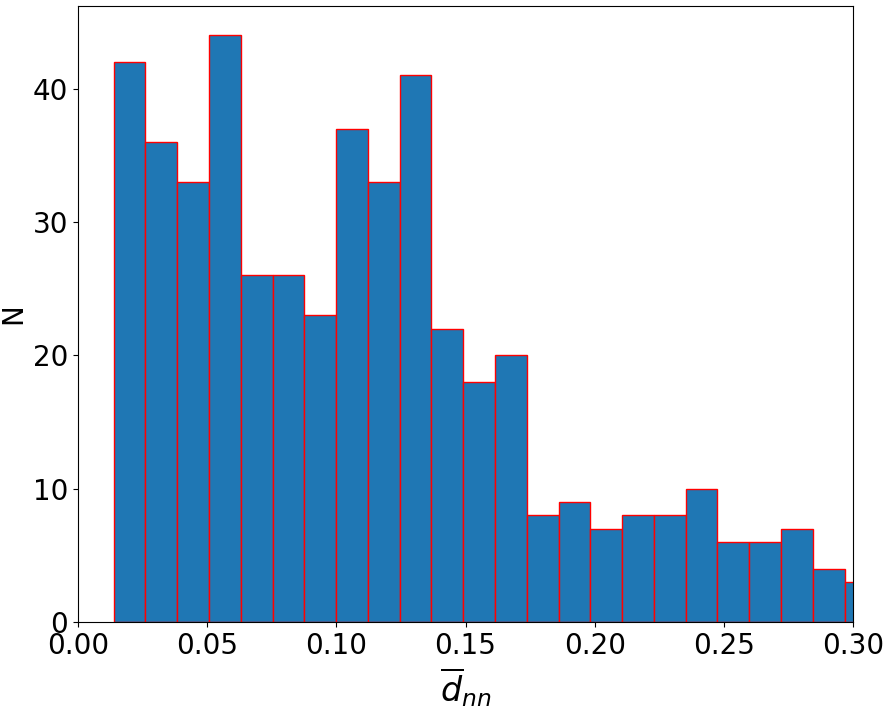}
\caption{}
\label{fig:knn1}
\end{subfigure}
\begin{subfigure}[b]{0.32\textwidth}
\centering
\includegraphics[width=\textwidth]{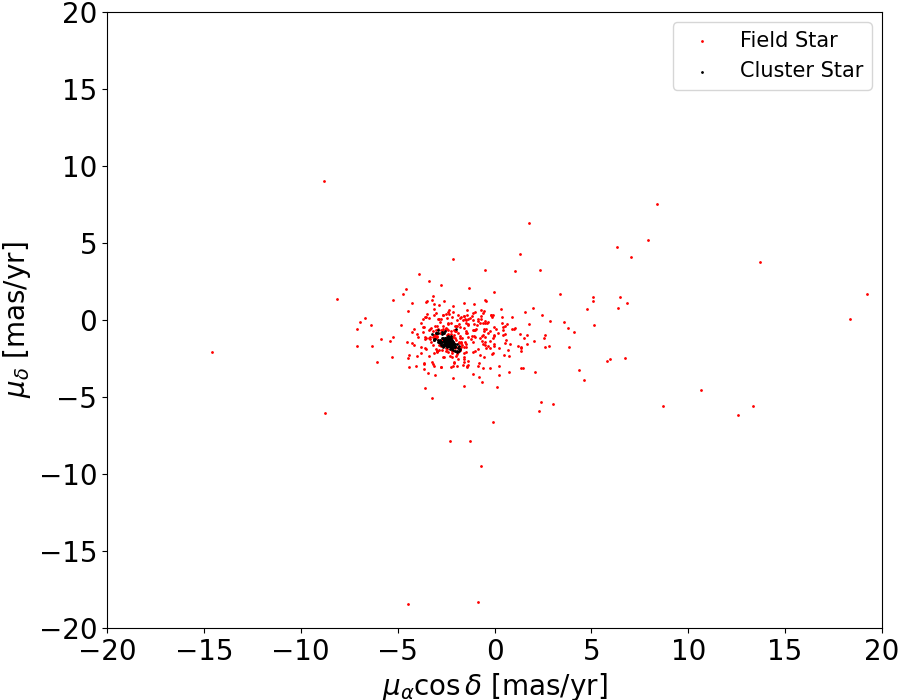}
\caption{}
\label{fig:knn2}
\end{subfigure}
\begin{subfigure}[b]{0.32\textwidth}
\centering
\includegraphics[width=\textwidth]{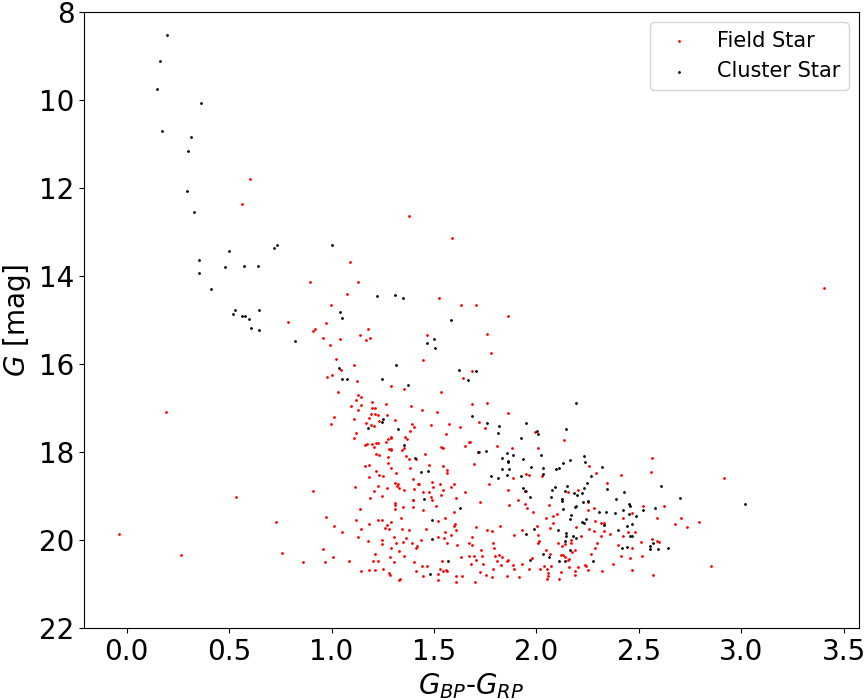}
\caption{}
\label{fig:knn3}
\end{subfigure}

\caption{(a) The distribution of $\overline{d}{\text{NN}}$ for the 588 stars identified within a search radius of 3 arcmin within the IC 1590 cluster. ;  (b) The distribution of proper motions for cluster stars ( black dots) selected based on $\overline{d}{\text{NN}}$ < 0.07 onto the dataset of 588 stars. ; (c) The color-magnitude diagram of the 172 cluster stars (black dots), selected based on $\overline{d}{\text{NN}}$ < 0.07 onto the dataset of 588 stars.}
\end{figure*}

\begin{figure*}
\centering
\begin{subfigure}[b]{0.32\textwidth}
\centering
\includegraphics[width=\textwidth]{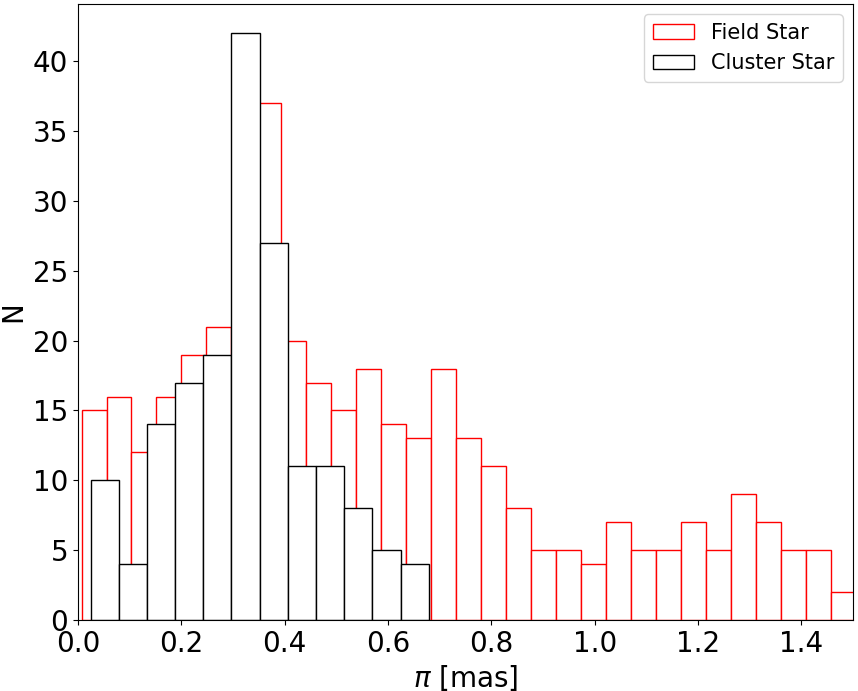}
\label{fig:knn4}
\end{subfigure}
\begin{subfigure}[b]{0.32\textwidth}
\centering
\includegraphics[width=\textwidth]{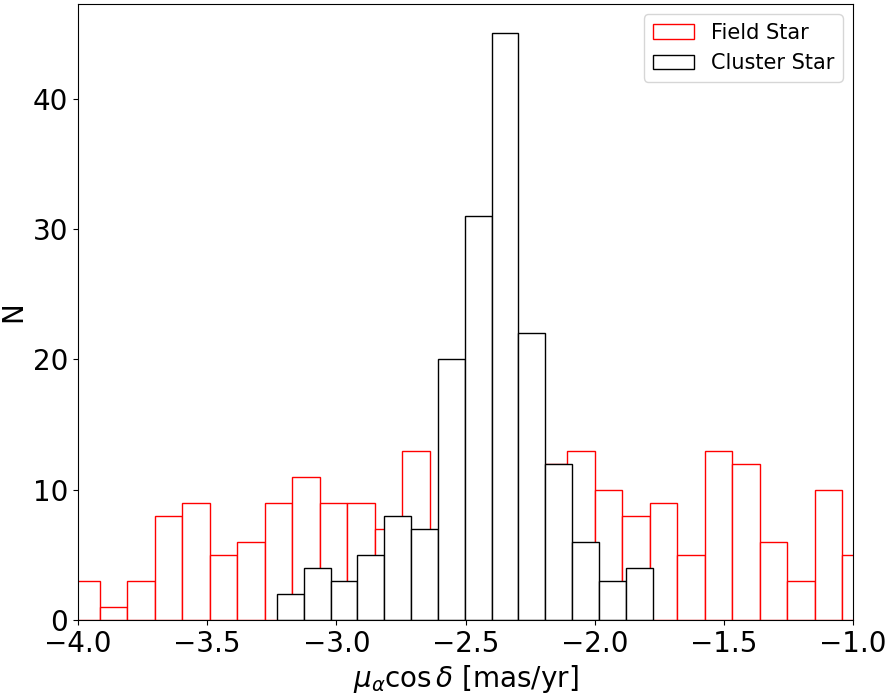}
\label{fig:knn5}
\end{subfigure}
\begin{subfigure}[b]{0.32\textwidth}
\centering
\includegraphics[width=\textwidth]{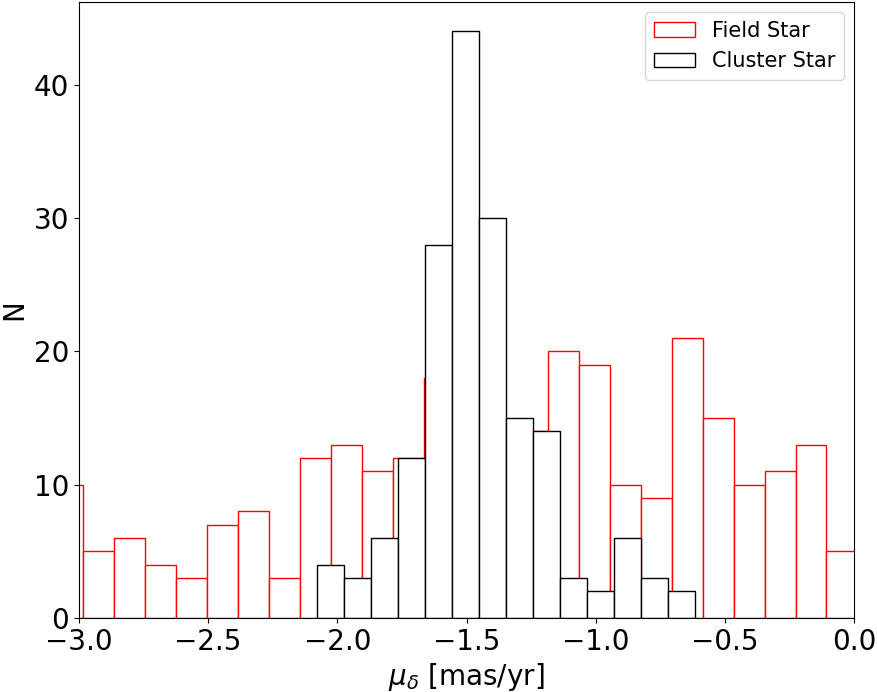}
\label{fig:knn6}
\end{subfigure}
\caption{The histogram plots of $\pi$,  $\mu_{\alpha}$$cos \delta$, $\mu_{\delta}$ of cluster stars (black), selected based on $\overline{d}{\text{NN}}$ < 0.07 onto the dataset of 588 stars.  are shown in the left, middle, and right figures, respectively.}
\label{fig:knndis}
\end{figure*}

The kNN technique for the outlier removal step is demonstrated using three astrometric data ($\pi$, $\mu{\alpha}$$\cos \delta$, $\mu_{\delta}$) of 588 stars obtained for IC 1590. The astrometric data of 588 stars is downloaded from the $Gaia$ Archive$^{\ref{fn:ex}}$ within a 3 arcmin search radius, using a search criterion of $\pi$  $\geq$ 0. The   distribution    of $\overline{d}_{\text{NN}}$ for  588   stars   is   shown  in Fig. \ref{fig:knn1}. A  cut-off value of $\overline{d}_{\text{NN}}$  = 0.07 is used to remove the outliers from the dataset and thus retain 172 stars. The selection of the threshold value (t) should be based on a relatively smaller value of $\overline{d}_{\text{NN}}$ to ensure that there are significantly more cluster stars compared to field stars \citep{10.1093/mnras/stac2116}. However, there is no predefined rule for choosing this threshold value. It's important to note that for cluster stars within a smaller search radius, the majority will be found in a densely populated region \citep{10.1093/mnras/stac2116}. In Fig. \ref{fig:knn2} and \ref{fig:knn3}, we overplotted these 172 stars (black) in the proper motion and color-magnitude diagram, respectively. Additionally, histograms in Fig. \ref{fig:knndis} illustrate the distribution of these 172 stars in the ($\pi$, $\mu_{\alpha}$$cos \delta$, $\mu_{\delta}$) parameter spaces as shown in the left, middle, and right figures, respectively. Table \ref{tab:knn} lists the range of parameters ($\pi$, $\mu_{\alpha}$$cos \delta$, $\mu_{\delta}$) resulting from this step for IC 1590. This range is then used to identify the members of the cluster within a 9 arcmin search radius from a larger dataset that was downloaded from the $Gaia$ Archive$^{\ref{fn:ex}}$. The parameter distributions exhibit distinct, sharp peaks that can be used for the identification of the distribution of cluster stars in these diagrams.

This step holds significant importance in the cluster membership analysis of open clusters using Gaussian Mixture Models (GMM). It plays a crucial role in eliminating a significant number of field stars from the sample. Moreover, it contributes to refining the parameters for the final cluster membership analysis of stars within a larger search radius from the downloaded sample \citep{10.1093/mnras/stac2116}. 
\begin{table}
\centering
\caption{ Selection of range of parameters ($\pi$,  $\mu_{\alpha}$$cos \delta$, $\mu_{\delta}$) for IC 1590 cluster.}
\begin{tabular}{cccc}
\toprule
Cut-off of $\overline{d}_{\text{NN}}$ & $\pi$ [mas] & $\mu_{\alpha}\cos\delta$ [mas yr$^{-1}$] & $\mu_{\delta}$ [mas yr$^{-1}$] \\ \hline
 0.07 & [ 0.0, 0.7 ] & [ -3.0, -2.0 ] & [ -2.0, -1.0 ] \\ 
\bottomrule
\end{tabular}
\label{tab:knn}
\end{table} 

\subsection{Gaussian Mixture Model on MD Distributions}
\label{sec:memgmm}
Mahalanobis distance (MD) \citep{mpc1927,article}  is  a  robust multi-variate distance metric,
that measures the distance between a point and a given distribution. It is a multidimensional generalization of the idea of measuring how many standard deviations away from the point is the mean of the distribution \citep{10.1093/mnras/stac2116}. To calculate the Mahalanobis Distance (MD) for each star, we used parallax ( $\pi$), proper motion in right ascension ($\mu_{\alpha}$$cos \delta$), and proper motion in declination($\mu_{\delta}$) relative to the center of the multivariate data. To ensure robust and comparable Mahalanobis Distance (MD) calculations across variables, we normalize the data using the covariance matrix. This normalization process breaks the correlation between different variables and standardizes them to the same scale, resulting in unitless distances \citep{10.1093/mnras/stac2116}. Thus
MD transforms multidimensional data into one dimension \citep{DEMAESSCHALCK20001}. The MD  of an observation  $\vec{x} =
(x_1,  x_2,  \ldots,  x_n)$  from  a set  of  observations  with  mean
$\vec{\mu} = (\mu_1,  \mu_2, \ldots, \mu_n)$ and  covariance matrix is
defined as \citep{article},
\begin{equation}
D_{M}(\vec{x}) = \sqrt{(x - \mu)^\text{T}\mathbf{\Sigma}^{-1}(x - \mu)}
\label{eq:md}
\end{equation}

\begin{table}
\centering
\caption{  Obtained mean value of parameters ($\pi$,  $\mu_{\alpha}$$cos \delta$, $\mu_{\delta}$) for IC 1590 cluster.}
\begin{tabular}{cccc}
\toprule
Radius [arcmin] & $\pi$ [mas] & $\mu_{\alpha}\cos\delta$ [mas/yr] & $\mu_{\delta}$ [mas/yr] \\ \hline
9 & 0.332 ± 0.05 & -2.37 ± 0.07 & -1.49 ± 0.08 \\
\bottomrule
\end{tabular}
\label{tab:gmm}
\end{table}

\begin{figure*}
\centering
\begin{subfigure}[b]{0.32\textwidth}
\centering
\includegraphics[width=\textwidth]{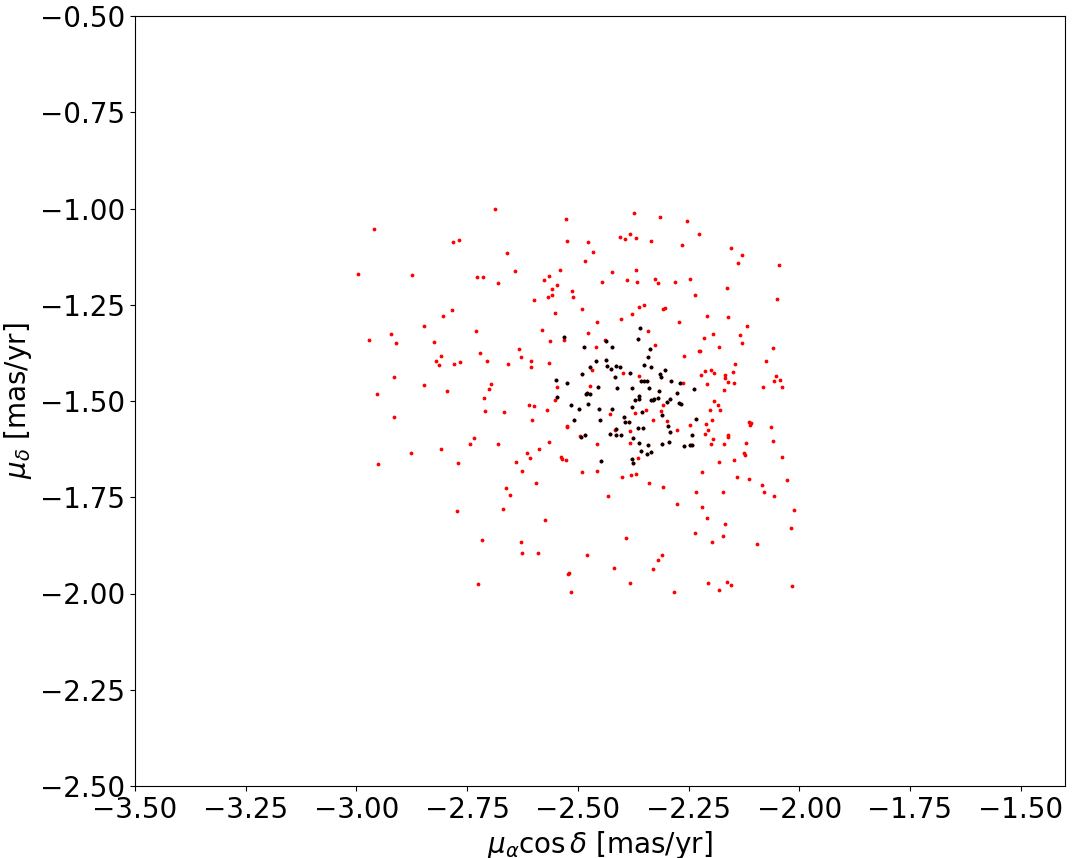}
\caption{}
\label{fig:gmm1}
\end{subfigure}
\begin{subfigure}[b]{0.32\textwidth}
\centering
\includegraphics[width=\textwidth]{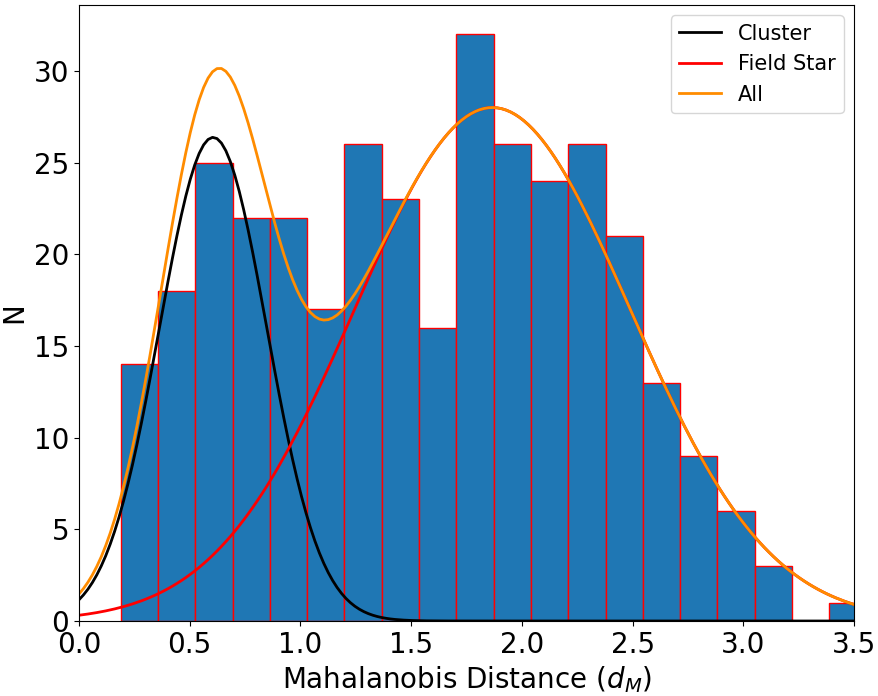}
\caption{}
\label{fig:gmm2}
\end{subfigure}
\caption{(a) The selected range of parameters used to obtain the proper motion plot of 344 stars from a search radius of 9 arcmin for the IC 1590 cluster, out of 4082 stars with a parallax of $\geq$ 0. ; (b) The distribution of Mahalanobis Distance (MD) for these 344 stars modeled using GMM with two-component and the resulting fits for the cluster, field stars, and both of them together.}
\label{fig:MD}
\end{figure*}

\begin{figure*}
\centering
\captionsetup{font=footnotesize}
\begin{subfigure}[b]{0.32\textwidth}
\centering
\includegraphics[width=\textwidth]{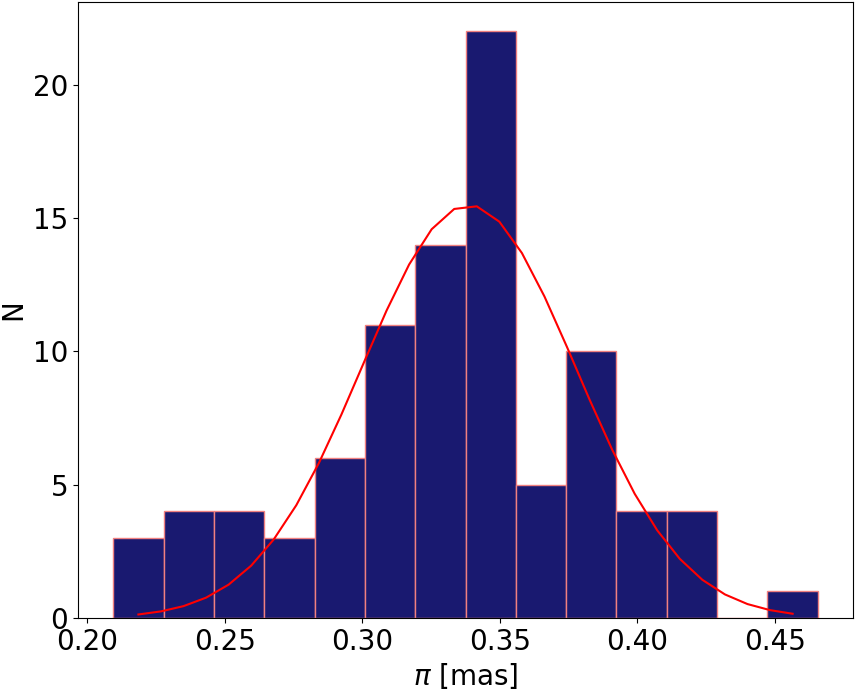}
\label{fig:gmm3}
\end{subfigure}
\begin{subfigure}[b]{0.32\textwidth}
\centering
\includegraphics[width=\textwidth]{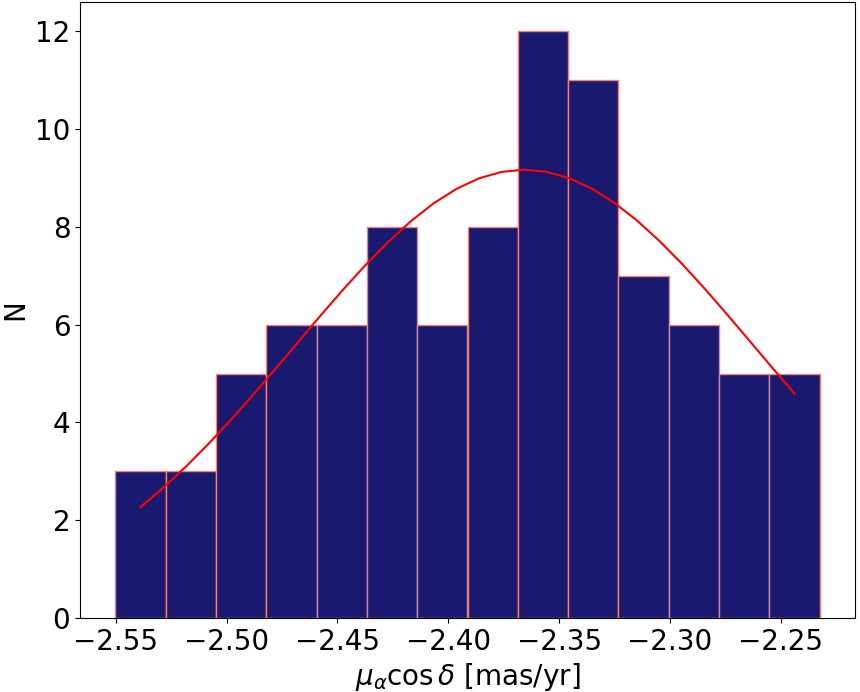}
\label{fig:gmm4}
\end{subfigure}
\begin{subfigure}[b]{0.32\textwidth}
\centering
\includegraphics[width=\textwidth]{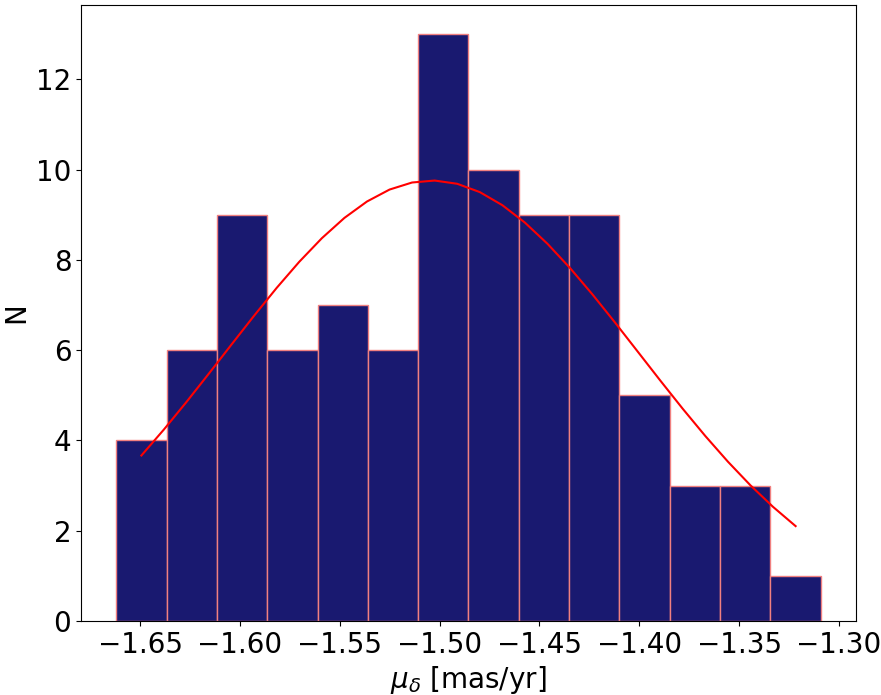}
\label{fig:gmm5}
\end{subfigure}
\caption{Gaussian fitted distribution of parallax ($\pi$) and individual proper motions ($\mu_{\alpha}\cos\delta$, $\mu_{\delta}$) to obtain the mean values of the cluster astrometric parameters for stars with membership probability > 0.50.}
\label{fig:par}
\end{figure*}

\begin{figure*}
\centering
\captionsetup{font=footnotesize}
\begin{subfigure}[b]{0.32\textwidth}
\centering
\includegraphics[width=\textwidth]{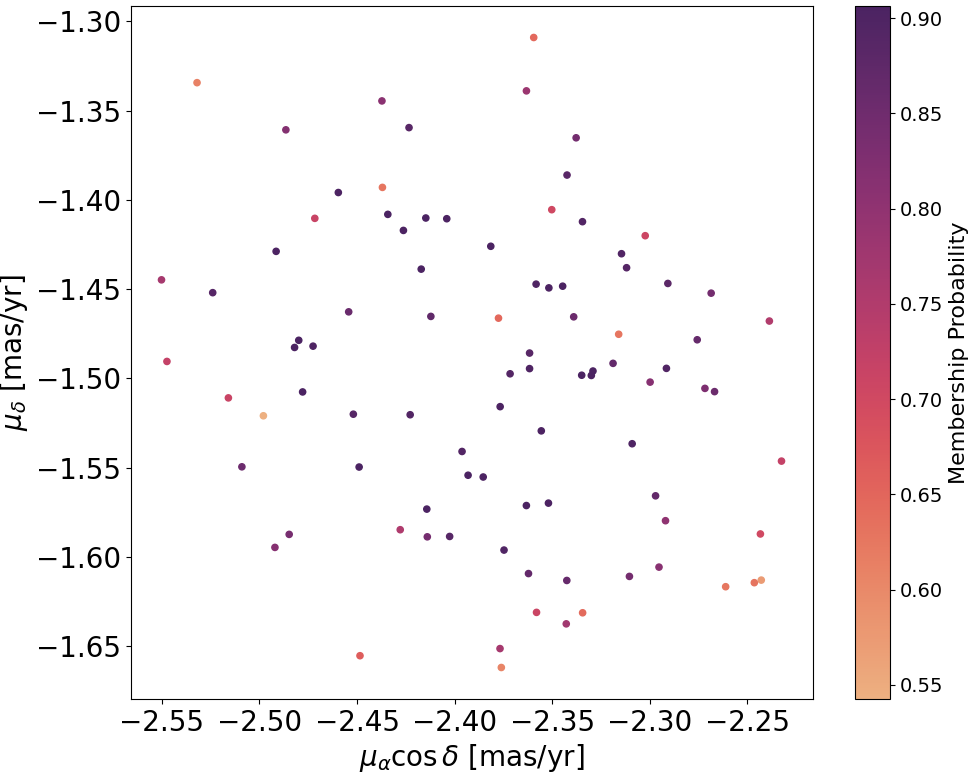}
\caption{}
\label{fig:gmm5}
\end{subfigure}
\begin{subfigure}[b]{0.32\textwidth}
\centering
\includegraphics[width=\textwidth]{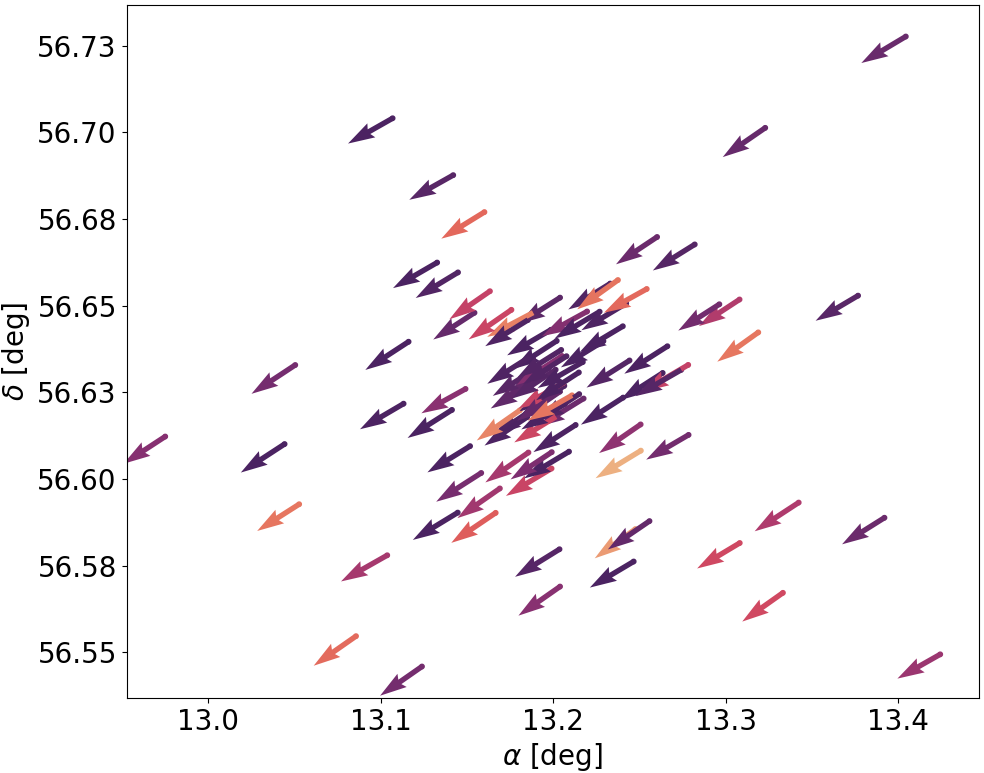}
\caption{}
\label{fig:gmm6}
\end{subfigure}
\begin{subfigure}[b]{0.32\textwidth}
\centering
\includegraphics[width=\textwidth]{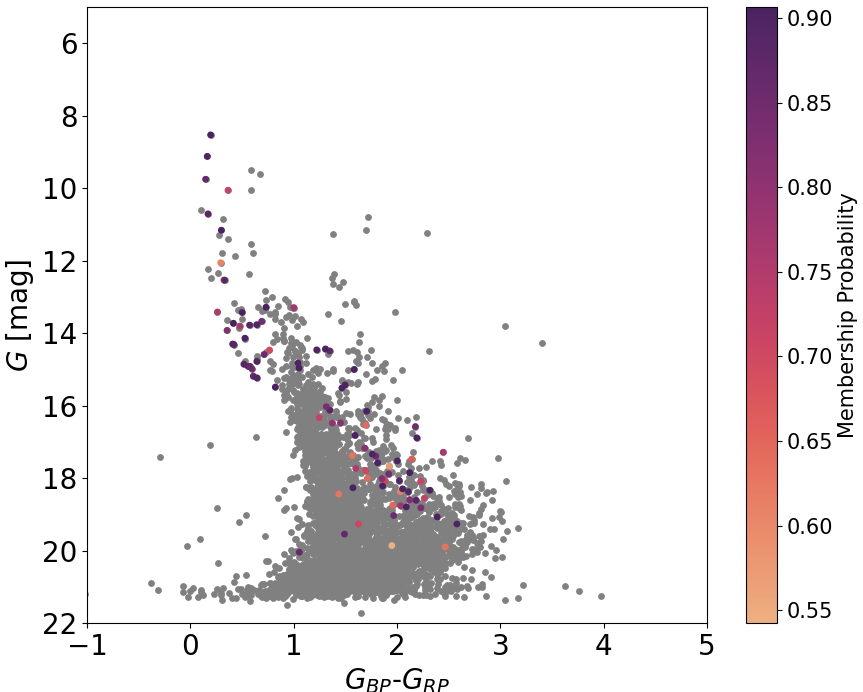}
\caption{}
\label{fig:gmm7}
\end{subfigure}
\caption{(a) The cluster members are showcased in the proper motion plane, with each member color-coded based on their membership probabilities. ; (b) The proper motion vector plot demonstrates the consistent direction of motion among member stars at their respective positions in the sky. ; (c) The color-magnitude diagram shows member stars plotted among all stars within a 9 arcmin radius, with each member color-coded based on their membership probabilities.}
\end{figure*}

A search within a 9 arcmin radius, considering stars with parallax $\geq$
0, yields a total of 4082 stars. Applying the selected range of parameters using the kNN technique in Section \ref{sec:knn} leaves us with 344 stars. Subsequently, these 344 stars are then subjected to MD calculation in the parameter space of ($\pi$, $\mu_{\alpha}$$\cos \delta$, $\mu_{\delta}$). The proper motion plot on  4082  stars and the distribution of the MD  of these 344 stars are
shown in Fig. \ref{fig:gmm1} and Fig. \ref{fig:gmm2}, respectively.

A Gaussian mixture model (GMM) is a statistical model that represents a probability density function by combining a finite number of Gaussian distributions, allowing it to approximate the given data distribution in parameter space \citep{MCLA2000,Deisenroth2020}.  It is one of the unsupervised machine
learning algorithms based on  Bayesian Decision Theory \citep{MCLA2000,10.5555/1403886}. In our analysis, the distribution of MD can be approximated by two Gaussian distributions: one corresponding to the cluster stars and the other to the field stars and variances of the probability distributions for cluster and field \citep{10.1093/mnras/stac2116}.\\ 
Let, $P_{c}(D_{M}|\mu_{c},\sigma_{c}^{2})$              and             the
$P_{f}(D_{M}|\mu_{f},\sigma_{f}^{2})$ represent the Gaussian probability
distributions of the cluster and the field stars, respectively. Then,
\begin{equation}
P(D_\text{M}|\mu,\sigma^{2}) = w_\text{c} P_\text{c}(D_\text{M}|\mu_\text{c},\sigma_\text{c}^{2}) + w_\text{f} P_\text{f}(D_\text{M}|\mu_\text{f},\sigma_\text{f}^{2})
\label{eq:gmm1}
\end{equation}
\&
\begin{equation}
w_\text{c} + w_\text{f} = 1
\label{eq:gmm2}
\end{equation}
where $w_{c}$, $\mu_{c}$, $\sigma_{c}$ represent the weights, means, and variances of the probability distribution corresponding to cluster stars, while $w_{f}$, $\mu_{f}$, $\sigma_{f}$ correspond to the same for the probability distribution related to field stars. The probability that a star with MD: $D_{M,i}$ will belong to a class ${k = (c, f)}$ is determined by the responsibility, called the membership probability \citep{Deisenroth2020}.
\begin{equation}
r_\text{ik} = \frac{w_\text{k} P_\text{k}(D_\text{M,i}|\mu_\text{k},\sigma_\text{k}^{2})} { w_\text{k} P_\text{k}(D_\text{M,i}|\mu_\text{k},\sigma_\text{k}^{2})}
\label{eq:gmm3}
\end{equation}
 Eq. \ref{eq:gmm1} is solved for the initial parameters ($w_{c}$, $\mu_{c}$, $\sigma_{c}$ and $w_{f}$, $\mu_{f}$, $\sigma_{f}$) using an unsupervised machine learning technique, called the expectation-maximization (EM) algorithm \citep{dempster1977maximum, fraley1998many, 10.5555/1403886,Deisenroth2020}. After carrying out the EM algorithm on initial parameters, the final values of parameters ($w_{k}$, $\mu_{k}$, $\sigma_{k}$) are obtained. Subsequently, the responsibility of all stars belonging to the cluster `c' (cluster membership probability) is calculated using,
\begin{equation}
r_\text{ic} = \frac{w_\text{c} P_\text{c}(D_\text{M,i}|\mu_\text{c},\sigma_\text{c}^{2})} { w_\text{k} P_\text{k}(D_\text{M,i}|\mu_\text{k},\sigma_\text{k}^{2})}
\label{eq:gmm4}
\end{equation}
Stars with $r_{ic}$ > 0.50 are considered to be members of the cluster `c'.\\
Using a Gaussian Mixture Model (GMM) with two components, we analyzed the Mahalanobis MD distribution of 344 stars. The GMM is employed to fit distributions representing the cluster, the field, and both components, as shown in Fig. \ref{fig:gmm2}. By setting a membership probability threshold of > 0.50, we identified 91 stars belonging to IC 1590.
We used Gaussian fits to analyze the distributions of parallax and proper motions ($\pi$, $\mu_{\alpha}$$\cos \delta$, $\mu_{\delta}$), to obtain the mean values of the cluster astrometric parameter for the member stars. The results of this analysis are presented in Fig. \ref{fig:par}, with the left, middle, and right figures, respectively. A Gaussian fit to the distribution results in mean values for the parameters ($\pi$, $\mu_{\alpha}$$\cos \delta$, $\mu_{\delta}$), listed
in Table \ref{tab:gmm}. The cluster stars are color-coded based on their 
membership probabilities in the proper motion plot as shown in Fig. \ref{fig:gmm5}. The ($G$ versus $G_{BP}$–$G_{RP}$)
color-magnitude diagram of these 91 cluster stars is shown in Fig. \ref{fig:gmm7}, with each member color-coded based on their membership probabilities. The directions of proper motions ($\mu_{\alpha}$$\cos \delta$, $\mu_{\delta}$) of the 91 cluster stars at their positions ($\alpha$, $\delta$) as shown in Fig \ref{fig:gmm6}. We can see that almost all the cluster members exhibit motion in the same
direction, indicating a  very effective determination of their membership
in the cluster.
Therefore, we conclude that the reliability of cluster membership analysis relies on the assumption that cluster members share the same kinematics.
\section{Multiwavelength Photometric Analysis}
\label{sec:multi}
\subsection{Photometric Analysis Based on Gaia Data}
\subsubsection{Astrophysical Parameters of IC 1590}
\label{sec:isodis}
\begin{figure}
\centering
\begin{subfigure}[b]{0.47\textwidth}
\centering
\includegraphics[width=\textwidth]{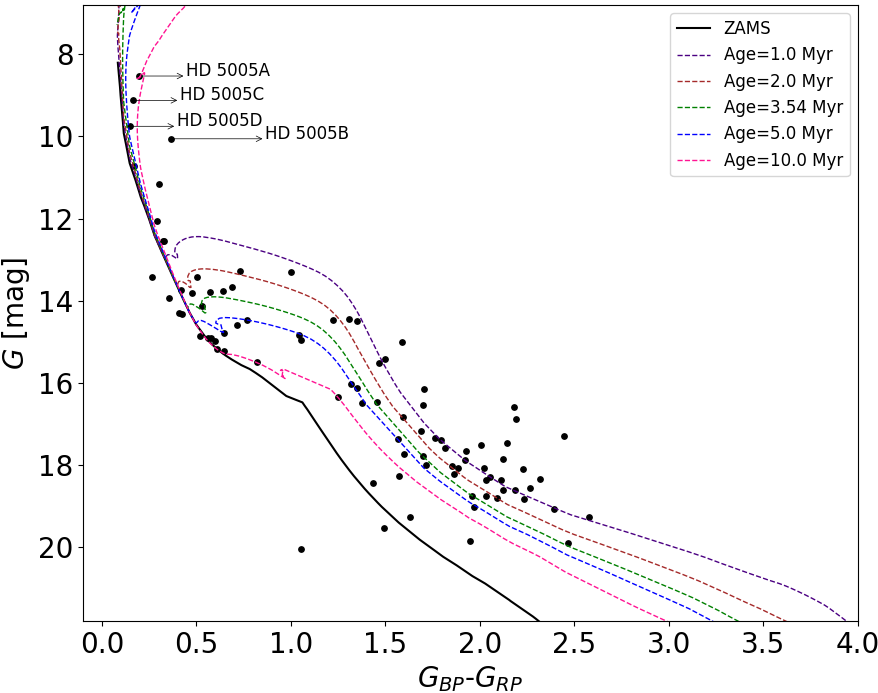}
\end{subfigure}
\caption{Color-magnitude diagram for cluster members of IC 1590 (black dots) with membership probability > 0.50. The black solid line is the
ZAMS plotted from MIST isochrone. The colored dotted lines are the PMS isochrones plotted from MIST isochrone for ages (1, 2, 3.54, 5, 10) Myr.}
\label{fig:gcmd}
\end{figure}

The $G$ versus ($G_{BP}$–$G_{RP}$) color-magnitude diagram for the cluster members of IC 1590 obtained in Section \ref{sec:memgmm} is shown in Fig. \ref{fig:gcmd}. In the cluster region, there is a moderately defined main sequence (MS) in the broad magnitude range, likely influenced by the variable reddening. This main sequence extends down to a magnitude of $G$ = 16. Furthermore, when examining the distribution of stars fainter than $G$ = 16 - 20 mag, it becomes apparent that these stars tend to shift toward the redder end of the main sequence, suggesting the presence of pre-main sequence (PMS) stars within the cluster region.

We used the Automated Stellar Cluster Analysis package (\href{https://github.com/asteca}{ASteCA}) \citep{2015A&A...576A...6P} which is a comprehensive suite of tools designed for the automated analysis of stellar clusters, aiming to determine their fundamental parameters. It includes an isochrone fitting procedure, utilizing synthetic clusters generated from theoretical isochrones, and optimizing the fit using a genetic algorithm. This allows \href{https://github.com/asteca}{ASteCA} to provide accurate estimations of a cluster's metallicity, age, extinction, and distance values, along with associated uncertainties \citep{2015A&A...576A...6P}. The best-fit process aims at estimating the distributions for a set of the cluster’s fundamental parameters. The process involves comparing the photometric diagram of the observed cluster with the diagrams of multiple synthetic clusters that have known parameter values. We estimated the best-fitted parameters for IC 1590 from isochrone fitting using \href{https://github.com/asteca}{ASteCA} based $Gaia$ EDR3 as distance $d$ $\sim$ 2.87 ± 0.02 Kpc, age $\sim$ 3.54 ± 0.05 Myr, metalicity $z$ $\sim$ 0.0212 ± 0.003, binarity value of $\sim$ 0.558 from \cite{2013ARA&A..51..269D} and extinction $A_v$ $\sim$ 1.252 ± 0.4 mag for a $R_v$ value of $\sim$ 3.322 ± 0.23. Since the most massive member of the cluster, IC 1590 is an O6.5 MS star \citep{1997AJ....113.2116G}, its maximum age should be of the order of the main-sequence lifetime of this massive star, that is 4.4 Myr \citep{1994A&AS..103...97M}. It is found that the IC 1590 has a maximum age of 3.5 Myr based on the location of pre-main sequence (PMS) stars on CMD \citep{1997AJ....113.2116G}. Our result aligns well with their findings.

We also used another powerful tool called the  MESA Isochrones \& Stellar Tracks \href{https://waps.cfa.harvard.edu/MIST/}{MIST} \citep{2016ApJ...823..102C}, which is based on the Modules for Experiments in Stellar Astrophysics (MESA) stellar evolution code. The MIST isochrone is a theoretical model that predicts the positions of stars on a CMD based on their age, metallicity, and distance. The MIST isochrone provides a theoretical framework that compares the observed CMD of a star cluster with predicted stellar positions based on different age, distance, and metallicity parameters. We fit the observed CMD to the MIST isochrone, using the best-fitted parameters obtained using (\href{https://github.com/asteca}{ASteCA}) is shown in Fig. \ref{fig:gcmd}. We identify the Trapezium-like system HD 5005, composed of four O-type stars HD 5005A (O4V(fc)), HD 5005B (O9.7II-III), HD 5005C (O8.5V(n)) and HD 5005D (O9.5V) from \citet{2011ApJS..193...24S} and marked in Fig. \ref{fig:gcmd}, using \href{https://aladin.cds.unistra.fr/}{$Aladin$ $sky$ $atlas$} \citep{2000A&AS..143...33B}, which is an interactive sky atlas allowing us to visualize digitized astronomical images or full surveys from different archives.

\subsubsection{Estimation of Luminosity and Mass Function of IC 1590}
\begin{figure}
\centering
\begin{subfigure}[b]{0.45\textwidth}
\centering
\includegraphics[width=\textwidth]{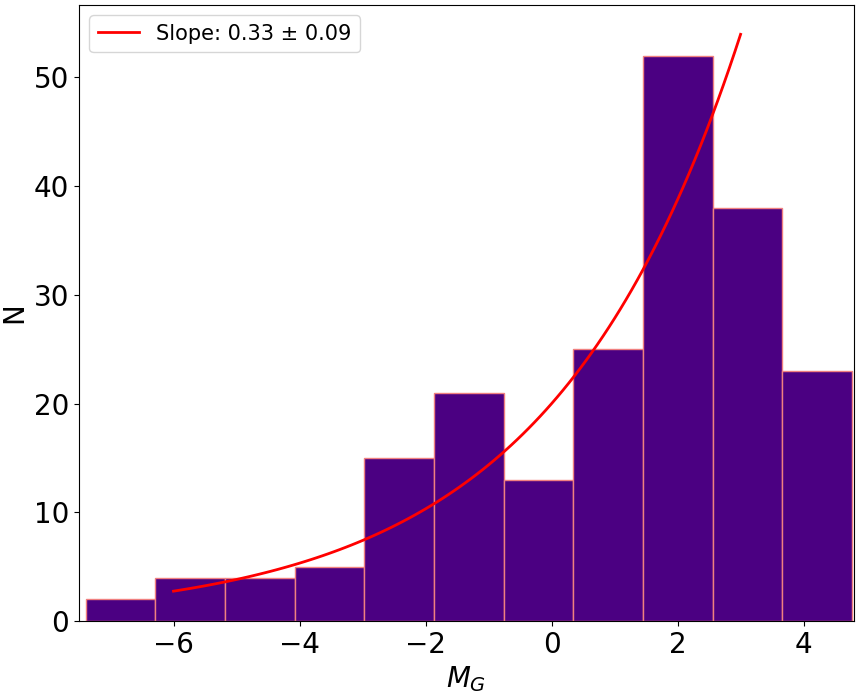}
\end{subfigure}
\caption{The $G$-band luminosity function based on selected stars ($P$ > 0.50) for IC 1590}
\label{fig:lf}
\end{figure}

The distribution of stellar brightness in a stellar cluster is known as its Luminosity Function (LF) \citep{2023PARep...1....1T}. To investigate the LF for IC 1590, we considered stars with membership probabilities $P$ > 0.50 within the 9 arcmin search radius obtained in the study. Additionally, we confined the magnitude range to 8 $\leq$ $G$ $\leq$ 20 mag to ensure the completeness of our cluster data. With these selection criteria, we perform the LF analysis for 91 stars. We used the apparent $G$ magnitudes of the selected cluster stars to calculate their absolute $M_G$ magnitudes using the distance modulus equation, given as $M_G$ = $G$ - 5×$log(d)$ + $A_G$, where $G$ represents the apparent magnitude and $d$ represents the isochrone distance, respectively. The value of $A_G$ is the extinction coefficient, which adjusts for the dimming of light due to interstellar dust. $A_G$ is the $G$-band extinction that is estimated for IC 1590 using the ratio $A_G$/$A_v$ = 1.13 \citep{Fukui_2019}. We plot as Fig. \ref{fig:lf} the LF histogram of IC 1590. It can be seen from Fig. \ref{fig:lf} that the absolute magnitudes of the selected stars fall within the range of -7.4 < $M_G$ < 4 mag. We perform linear regression to estimate the slope of the LF, which yields the slope of the $G$-band LF = 0.33 ± 0.09. In Fig.  \ref{fig:lf}, the luminosity function peaks between
$M_G$ $\sim$ 2 and 3 mag, and below 3 mag the luminosity function steadily declines.

The initial mass function (IMF) is defined as the distribution of initial masses within a population of stars \citep{2015A&A...576A...6P}. Since the properties and evolution of a star are intimately linked to its mass, the IMF serves as a crucial tool for studying large populations of stars. So, young clusters are valuable to investigate the IMF, since they are too young to lose a significant number of members either by dynamical process or stellar evolution. The IMF, is often referred to as a probability density function (PDF) that describes the probability of a star that has a certain mass during its formation \citep{2003PASP..115..763C}. For a population of $N$ stars with masses $m_i$ and a total mass of $M_T$, the IMF $\xi(m)$ is given by,
\begin{align}
IMF \rightarrow \xi(m) = \frac{dn}{dm} \rightarrow dn = \xi(m) \, dm\\
M_T = \sum_{i=1}^{N} m_i \rightarrow M_T = C \int_{m_l}^{m_h} m(n) \, dn \nonumber\\
 =  C \int_{m_l}^{m_h} m \xi(m) \, dm
\end{align}
where, $m_l$ and $m_h$ are the mass limits for the IMF and $C$ is a normalization constant. Setting the total mass to unity, $M_T$ = 1M\sun, allows us to obtain the normalization constant $C$ and treat the normalized IMF as a PDF:
\begin{equation}
M_T = 1M\sun \to C = \frac{1}{\int_{m_l}^{m_h} m\xi (m)  dn}
\label{eq:gmm4}
\end{equation}
and thus the normalized IMF can be written as:
\begin{equation}
PDF(m) = \xi (m)_{norm} = C \xi (m)
\label{eq:gmm4}
\end{equation}
Once the PDF is generated, every time a new synthetic cluster is created and samples several masses randomly from it, following the probability distribution given by the PDF. This process continues until the total mass fixed by the total-mass parameter is attained, providing a distribution of masses probabilistically sampled from a certain IMF with their sum up to the total cluster mass \citep{2015A&A...576A...6P}. The synthetic star clusters (SSC) are generated by \href{https://github.com/asteca}{ASteCA} with given metallicity, age, extinction, distance, mass, and binarity values, mimicking those of the input observed cluster.

\begin{figure}
\centering
\begin{subfigure}[b]{0.45\textwidth}
\centering
\includegraphics[width=\textwidth]{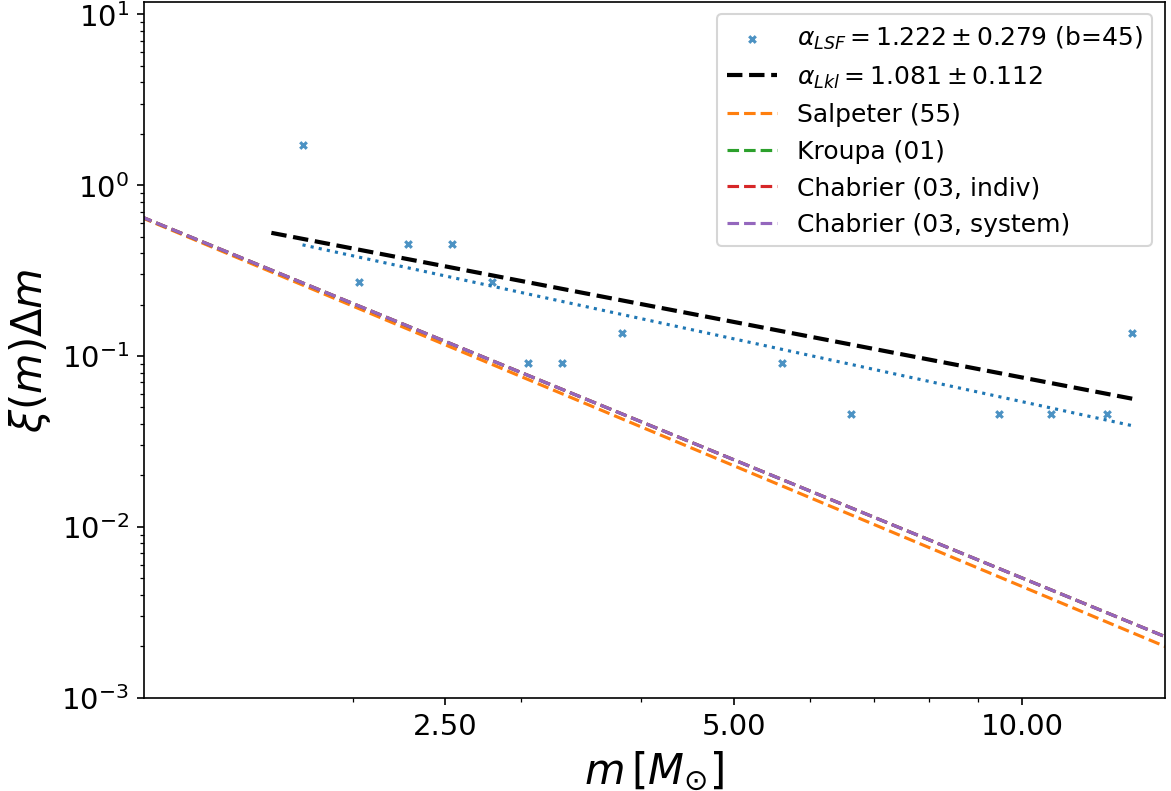}
\end{subfigure}
\caption{The IMF of the synthetic cluster when binary stars are considered as single stars.}
\label{fig:single}
\end{figure}

\begin{figure}
\centering
\begin{subfigure}[b]{0.45\textwidth}
\centering
\includegraphics[width=\textwidth]{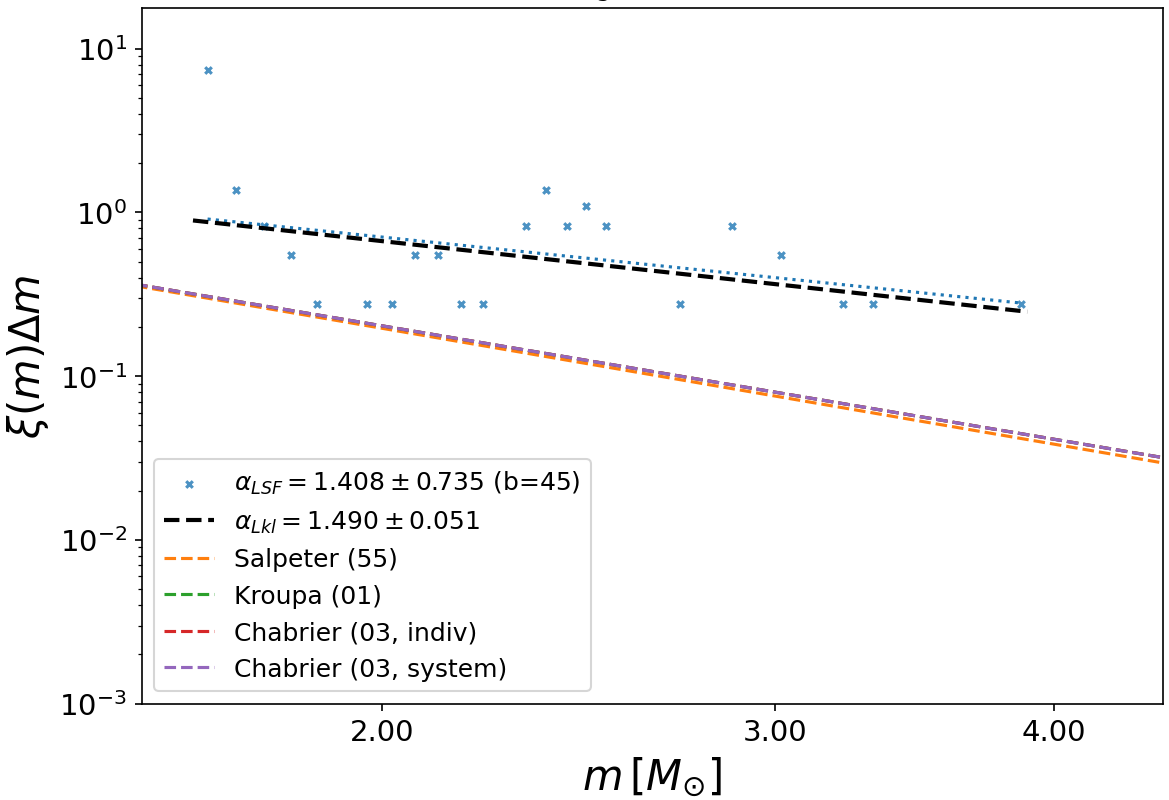}
\end{subfigure}
\caption{The IMF of the synthetic cluster taking into account the binarity value of $\sim$ 0.558 estimated by ASteCA.}
\label{fig:binary}
\end{figure}

We estimate the IMF for the observed cluster IC 1590 using \href{https://github.com/asteca/imf}{ASteCA IMF}, given its masses and uncertainties are estimated from SSC generated by  \href{https://github.com/asteca}{ASteCA}. The method employed here was originally developed by \cite{2013MNRAS.434.3236K}. Binary stars, whether they are primordial or dynamically formed during close encounters between single stars, can have an impact on the mass function. So, the characterizing the binary fraction is very important. The likelihood analysis is performed on a selected mass range of (1, 100) M$\sun$ for estimation of the slope of IMF and we found the slope of IMF is $\alpha$ = 1.081 ± 0.112 for a total mass of $M_T$ = 255 M$\sun$ when binary stars are considered as single stars as shown in Fig. \ref{fig:single} by the dotted black line and $\alpha$ = 1.490 ± 0.051 for a total mass of $M_T$ = 152 M$\sun$ taking into account the binarity value of $\sim$ 0.558 estimated by \href{https://github.com/asteca}{ASteCA} from \cite{2013ARA&A..51..269D} as shown in Fig. \ref{fig:binary} by the dotted black line. We also performed least-square fitting on the selected mass range using a number of bins = 45 and got the slope $\alpha$ = 1.22 ± 0.279 when binary stars are regarded as single stars as shown in Fig. \ref{fig:single} by the dotted blue line and $\alpha$ = 1.408 ± 0.735 taking into account the binarity value of $\sim$ 0.558 as shown in Fig. \ref{fig:binary} by the dotted blue line. The remaining colored dashed lines are the slopes of several theoretical IMFs \cite{1955ApJ...121..161S}, \cite{2001MNRAS.322..231K}, and \cite{2003PASP..115..763C}, for comparison.

\subsection{Photometric Analysis based on 2MASS NIR Data}
\subsubsection{Identification of YSOs}
\label{sec:yso}
\begin{figure}
\centering
\begin{subfigure}[b]{0.48\textwidth}
\centering
\includegraphics[width=\textwidth]{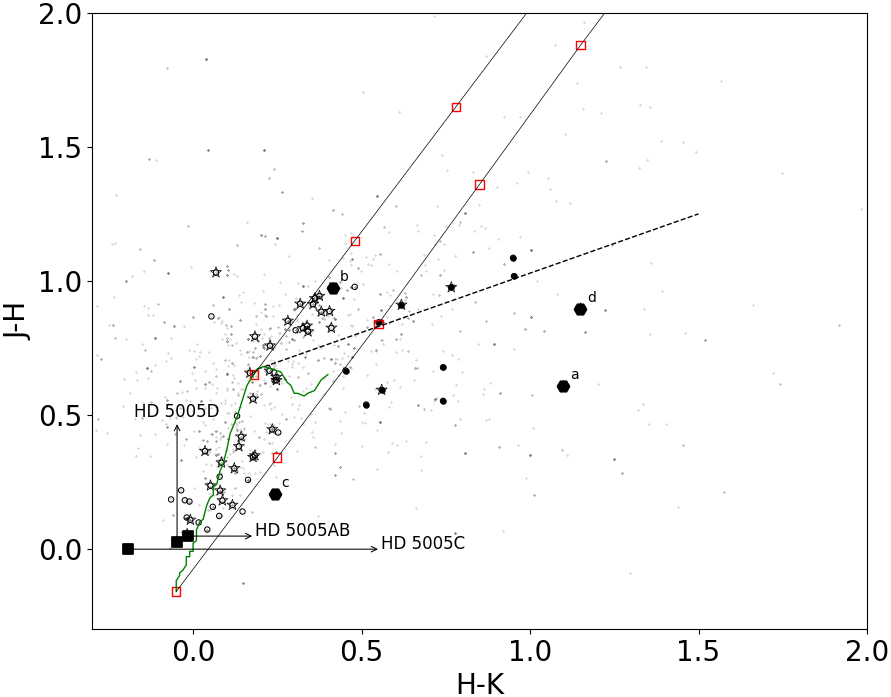}
\end{subfigure}
\caption{NIR color-color diagram of the detected 2MASS counterparts (open circles), X-ray counterparts  (star symbols), O-type stars (black-filled square symbols), NIR-excess YSOs (black-filled circles), and H$_{\alpha}$ sources  (black-filled hexagon symbols) in the cluster region.}
\label{fig:tcd}
\end{figure}

Young stellar objects (YSOs) can be classified based on their evolutionary stage, and their positions on the $(J - H)$ versus $(H - K)$ color-color diagrams (CCs) are indicative of this. Different types of YSOs, such as protostellar-like objects, T Tauri stars, Herbig Ae/Be stars, and classical Be stars, occupy distinct regions on NIR CCs \citep{2012PASJ...64..107S}.\\
The NIR $JHK_s$ magnitudes of the stars within a circle of radius 9 arcmin were obtained from the 2MASS point source catalogue \citep{2003yCat.2246....0C}, yields a total of 2144 2MASS point sources in this cluster region. A total of 68 2MASS counterparts were found within a search radius of 1$''$. We took a list of 975 X-ray sources in this from The 3rd MSFRs Omnibus X-ray Catalog (MOXC3) \citep{2020yCat..22440028T} in this region. We found a total of 42 X-ray counterparts within a search radius of 1$''$. We also took a list of 1593 stars in this from The SOS. VII. UBVI photometry of open cluster IC 1590 \citep{2021yCat..51620140K} in this region. We found a total of 4
H$_{\alpha}$ counterparts within a searching radius of 1$''$. The $JHK_s$ colors were transformed from the 2MASS system to the Koornneef system using the relations given by \cite{2001AJ....121.2851C},
\begin{align}
(J-H)_\text{2MASS} &= (1.024 \pm 0.024)(J-H)_\text{Kf} + (-0.045 \pm 0.006) \\
(J-K_\text{s})_\text{2MASS} &= (0.970 \pm 0.015)(J-K_\text{s})_\text{Kf} + (-0.017 \pm 0.005) \\
(H-K_\text{s})_\text{2MASS} &= (0.792 \pm 0.056)(H-K_\text{s})_\text{Kf} + (0.027 \pm 0.005)
\end{align}
We show the $(J - H)$ versus $(H - K)$ color-color diagram for all the selected sources (small dots) from 2MASS NIR measurements in Fig. \ref{fig:tcd}. We assume that the \cite{1985ApJ...288..618R} reddening law can be applied to the IC 1590 cluster region and represents a reasonable approximation of the NIR extinction that is caused by the associated molecular dust and material. We have plotted as green solid lines in Fig. \ref{fig:tcd}, the unreddened main-sequence (MS) stars \citep{1983A&AS...51..489K}. From the extreme points of this green curve, we have plotted two black solid lines parallel to the \citet{1985ApJ...288..618R} interstellar reddening vector, and on these black solid lines, we have marked points with red open squares at an interval of $A_v$ = 5 mag \citep{2007MNRAS.379.1237M}. The area between these two lines corresponds to the reddening zone for normal stars, which are considered to be either field stars (MS, stars, giants) or Class III and Class II sources with small NIR-
excesses \citep{1992ApJ...393..278L,2012PASJ...64..107S}. The region that is to the right-hand side of the reddening band is known as the NIR excess region \citep{1992ApJ...393..278L} and corresponds to the location of YSOs, mostly classical T Tauri stars (CTTSs) with large NIR-excesses \citep{2007MNRAS.379.1237M}.
We have plotted as a black dashed line extending from the extreme points of this MS curve, the dereddened classical T Tauri locus taken from \citet{1997AJ....114..288M} that is converted from the CIT system using the transformation relations given by \citet{2001AJ....121.2851C}. Stars lying below the CTTSs locus and in the NIR excess region in Fig. \ref{fig:tcd} can be explained by the circumstellar disc models of  \citet{1992ApJ...393..278L} and their standard models account for a range of disc inclinations, radial temperature profiles, and stellar surface temperatures. However, some young stars such as naked T-Tauri stars, post-T Tauri stars, and some class-I sources do not show any near-infrared excess, and they will appear between the two reddening lines on such a diagram \citep{2007MNRAS.379.1237M}.

Out of the 68 2MASS point sources shown by open circles, 13 sources are situated in the NIR excess region, illustrated in Fig. \ref{fig:tcd} using black-filled circles. Grey dots represent all the 2MASS point sources detected in the cluster region. The locations of the Trapezium-like system HD 5005, which is composed of four O-type stars named HD 5005A, HD 5005B, HD 5005C, and HD 5005D, have been identified using black-filled square symbols in Fig. \ref{fig:tcd}. However, it is worth noting that the two brightest stars in IC 1590, namely HD 5005A and HD 5005B, are too bright to be measured reliably \citep{2021AJ....162..140K}. So, they were not resolved in the 2MASS point source catalogue \citep{2003yCat.2246....0C}. So, we consider it as HD 5005AB. The identified X-ray sources are shown as star symbols. The identified H$_{\alpha}$ sources are shown as black-filled hexagon [a, b, c, d] symbols in Fig. \ref{fig:tcd}. \citet{2010BASI...38...35M} reported that the H$_{\alpha}$ source[c] a CBe star is spectral-type B2, H$_{\alpha}$ source[a] an HBe star is spectral-type B8.5 and H$_{\alpha}$ source[d] an HAe star is A0, and found that nebulosity associated with [a] and [d].

\begin{figure}
\centering
\begin{subfigure}[b]{0.45\textwidth}
\centering
\includegraphics[width=\textwidth]{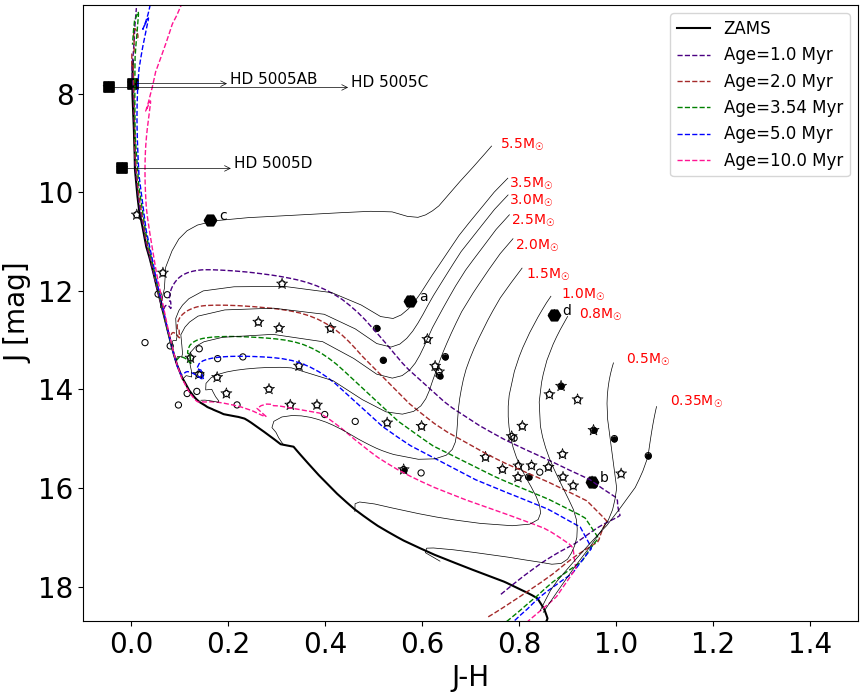}
\end{subfigure}
\caption{The NIR $J$ versus $(J-H)$ CMD for the for 2MASS counterparts (open circles), X-ray counterparts  (star symbols), O-type stars (black-filled square symbols), NIR-excess YSOs (black-filled circles) and H$_{\alpha}$ sources  (black-filled hexagon symbols) in the IC 1590 region.}
\label{fig:ccm}
\end{figure}

\subsubsection{Age and Mass Estimation of YSOs}
The age and mass of young stellar objects (YSOs) can be estimated by comparing their location with the theoretical isochrones. In general, the right side of the main sequence (MS) on the color-magnitude diagram (CMD) for a cluster is occupied by field giants. Interestingly, this is the same region that could also be inhabited by pre-main sequence (PMS) stars if they are part of the cluster population. As discussed in Section \ref{sec:yso}, we identified some stars with near-infrared (NIR) excess. If these stars are indeed PMS stars, they should appear to the right of the main sequence in the CMD. To put it simply, the presence of NIR excess and their position in the NIR CCs can serve as a useful means to distinguish PMS stars from the field stars found on the right side of the CMD \citep{2006MNRAS.370..743S}.

We represent the $J$ versus $(J-H)$ CMD in Fig. \ref{fig:ccm}. The various symbols represent 2MASS counterparts (open circles), X-ray counterparts  (star symbols), O-type stars (black-filled square symbols), NIR-excess YSOs (black-filled circles), and H$_{\alpha}$ sources  (black-filled circle symbols) are shown in the Fig. \ref{fig:ccm}. We fit the ZAMS using the MIST isochrone \citep{2016ApJ...823..102C} for total iron abundance, $[Fe/H]$ = 0.047 obtained by converting the $z$ = 0.0212 described by \citet{2015A&A...576A...6P}, distance $d$ = 2.87 Kpc and visual extinction $A_v$ = 1.252 as obtained in Section \ref{sec:isodis} shown by black solid line in Fig. \ref{fig:ccm}. To estimate the ages and masses of the YSOs, we plotted  few PMS isochrones for ages 1, 2, 3.54, 5, and 10 Myr  are shown by colored dashed lines, while the isochrones for masses 0.35 - 5.5 M$\sun$ are shown by grey solid lines in Fig. \ref{fig:ccm}.\\
From the NIR $J$ versus $(J-H)$ CMD, we found that the identified NIR-excess YSOs are probably the pre-main sequence (PMS) stars of ages $\leq$ 2 Myr and masses $\sim$ 0.35 - 5.5 M\sun.  Most of the PMS stars are undergoing evolution along Hayashi tracks or in the Kelvin–Helmholtz contraction phase \citep{2021AJ....162..140K}.

\subsubsection{Extinction Map}
Dust extinction is the most reliable tracer of the gas distribution in the interstellar
medium and for exploring the properties of dust clouds, but measuring extinction is limited by the systematic uncertainties involved in estimating the intrinsic colors of background stars \citep{2017A&A...601A.137M}. The precision of dust extinction measurement is fundamentally linked to the intrinsic color of background stars. The challenges lie in accurately estimating the uncertainties associated with determining this intrinsic color,
which has in turn consistently impeded the accuracy of dust extinction measurements.

\citet{1994ApJ...429..694L} introduced a method for measuring extinction in the dark cloud IC 5146 using $JHK$ photometric data from infrared imaging surveys. Subsequently, \citet{Alves_1998} modified this technique into the Near Infrared Color Excess (NICE) method, which used deep NIR data to determine color excess in dark clouds. \citet{2001A&A...377.1023L} improved the NICE technique, renaming it NICER, by incorporating a nearby extinction-free control field to assess the distribution of intrinsic colors better, resulting in a more robust approach, assuming a well-determined reddening law is applied. However, challenges such as foreground contamination and bias in column density due to unresolved cloud substructures persisted. \citet{2009A&A...493..735L} came up with the NICEST technique, which effectively mitigated biases introduced by sub-pixel structures and contamination from foreground stars.

\begin{figure}
\centering 
\begin{subfigure}[b]{0.485\textwidth}
\centering
\includegraphics[width=\textwidth]{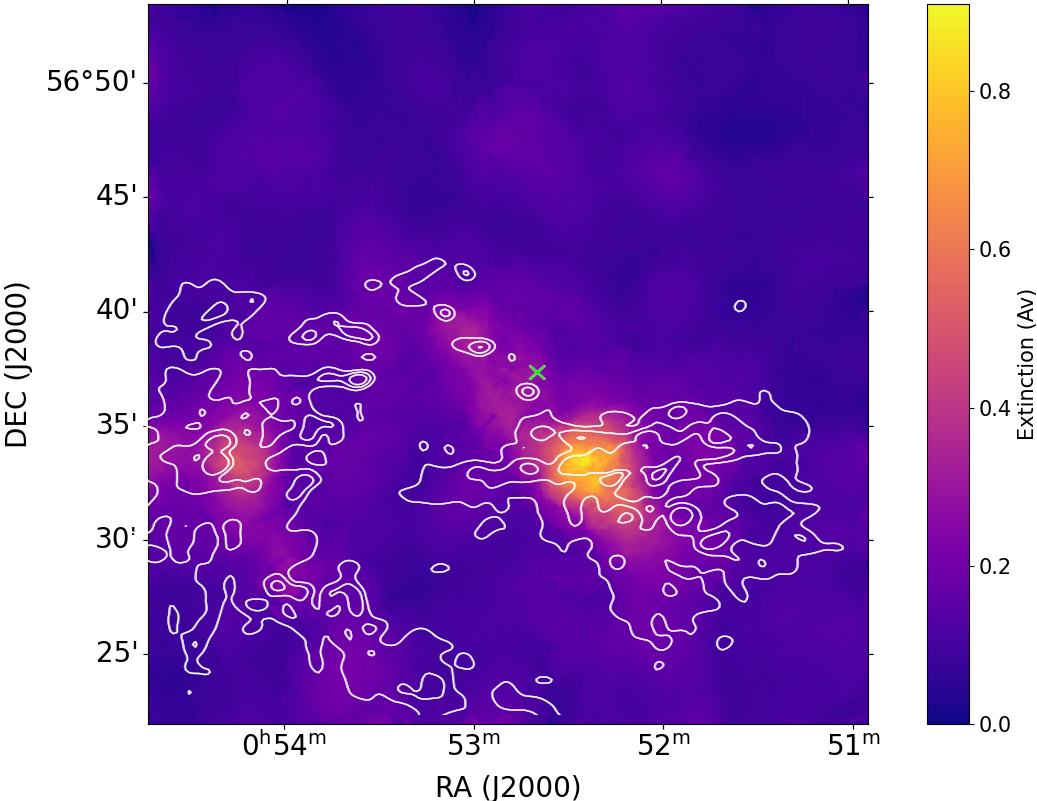}
\end{subfigure}
\caption{PNICER  visual extinction map of NGC 281 region with a FoV of  30$'$$\times$30$'$. The visual extinction in magnitudes of Av is represented in colors. The center of the cluster (00:52:49.2, +56:37:44) is shown by the green 'X' mark. The contours extracted from $Herschel$ SPIRE 500 $\mu$$m$ dust continuum emissions are plotted over the extinction map.}
\label{fig:emap}
\end{figure}

\citet{2017A&A...601A.137M} introduced the \href{http://smeingast.github.io/PNICER/ }{PNICER} technique, designed to estimate the visual extinction towards individual point sources, which uses unsupervised machine learning techniques to calculate extinction towards individual point sources. \href{http://smeingast.github.io/PNICER/ }{PNICER} technique uses a probability density function (PDF) determined from Gaussian mixture models (GMM) fitting along the extinction vector to determine extinction towards individual sources, using control field observations that are free from extinction effects \citep{2017A&A...601A.137M}. This approach provides superior accuracy in calculating extinction compared to NICER and NICEST, as it minimizes the variance in intrinsic color calculations by considering an extinction-free region in proximity to the target field. \href{http://smeingast.github.io/PNICER/ }{PNICER} converts the determined intrinsic parameter PDF into the visual extinction by comparing the distribution to the observed parameters while relying on a defined extinction law given by \citet{2005ApJ...619..931I}. In this way, it becomes possible to describe the visual extinction for the observed sources with probability densities, rather than a single value \citep{2017A&A...601A.137M}.
In this study, we utilized the \href{http://smeingast.github.io/PNICER/ }{PNICER} technique to determine the visual extinction for individual sources within the NGC 281 regions. The method relies on NIR photometric data, obtained from the 2MASS point source catalogue \citep{2003yCat.2246....0C}. We only considered data points with photometric uncertainties available for all three filters ($JHK_s$) when measuring extinction. The region containing IC 1590 is shown for a field of view of 30$'$$\times$30$'$ in Fig. \ref{fig:emap}. The visual extinction map for the NGC 281 region is shown in Fig. \ref{fig:emap}. The extinction values range from approximately 0-1.45 mag, with the highest extinction of 1.45 mag observed near the NGC 281 West region.

Moreover, due to the availability of $Herschel$ SPIRE 500 $\mu$$m$ dust continuum emissions data \citep{2010A&A...518L...3G}, it is possible to compare the density structures
with the dust structure revealed by the extinction in the NGC 281 region. The contours extracted from the 500 $\mu$$m$ dust continuum emissions map of $Herschel$ SPIRE  are overplotted on the extinction map of the NGC 281 region as shown in Fig. \ref{fig:emap}. The dust structure revealed by the extinction appears to be consistent with the contours in the NGC 281 region.

\section{Distance Estimation of IC 1590 based on parallax}
\label{sec:dis}
It is very important to know how to estimate distances (and their uncertainties)
from parallaxes \citep{2015PASP..127..994B}. While the relationship between parallax and distance is straightforward, the inversion of parallax to calculate distance is valid only under ideal conditions i.e. without any measurement uncertainties. As we always have measurement errors, determining the distance from parallax values becomes a complex inference problem that requires careful consideration of these uncertainties.
In this study, we adopt the exponentially decreasing space density (EDSD) prior in
distance is given by \citep{2018AJ....156...58B},
\begin{align}
P(r|L) = \begin{cases}
\frac{1}{2L^{3}}r^{2}\exp\left(-\frac{r}{L}\right) & \text{if } r > 0 \\
0 & \text{otherwise}
\end{cases}
\label{eq:bj}
\end{align}
where $L$ > 0 is a length scale. The prior exhibits a single mode at $2L$ and $L$ varies as a
function of galactic longitude ($l$), and galactic latitude ($b$), according to a model, to anticipated changes in the distribution of stellar distances within the $Gaia$-observed Galaxy \citep{2018AJ....156...58B}.

For a star at a true distance of $r$, its parallax is $1/r$, which is unknown, but the measured parallax $\overline{\omega}$ is a noisy measurement of $1/r$ \citep{2015PASP..127..994B}.  Taking into consideration the Gaussian likelihood in the parallax $\overline{\omega}$ with standard deviation $\sigma_{\overline{\omega}}$, and the EDSD prior from Eq. \ref{eq:bj}, we can obtain the unnormalized posterior over the distance to a source  \citep{2018AJ....156...58B},
\begin{align}
& P^{*}(r|\overline{\omega},\sigma_{\overline{\omega}}, L_{\text{sph}}(l,b)) \nonumber\\
&= \begin{cases}
    r^{2}\exp\left[-\frac{r}{L_{\text{sph}}(l,b)}-\frac{1}{2\sigma_{\overline{\omega}}^2}(\overline{\omega}-\overline{\omega}_{zp}-\frac{1}{r})^2\right] & \text{if } r > 0 \\
    0 & \text{otherwise}
  \end{cases}
\label{eq:bj2}
\end{align}
This function is the measurement model or likelihood \citep{2018AJ....156...58B}. It gives the probability density function (PDF) that is, probability per unit parallax for any $\overline{\omega}$, given values of $r$ and $\sigma_{\overline{\omega}}$. $\omega_{zp}$ is the global parallax zero point, determined from Gaia’s observations of quasars to be -0.029 mas \citep{2018A&A...616A...2L}. For physical values of its three
parameters: finite $\overline{\omega}$, positive  $\sigma_{\overline{\omega}}$, positive $L_{sph}$, it is always a
proper (i.e., normalizable) density function \citep{2018AJ....156...58B}.

We utilized the astrometry data from $Gaia$ DR2 and the length scale model to estimate distances. Due to the limited precision of fractional values, the accuracy of parallax measurements is constrained, particularly for stars located at greater distances. \cite{2018AJ....156...58B} opt to estimate distances without making specific assumptions about the characteristics of individual stars or the extent of extinction affecting them. It's important to acknowledge that even though we calculate distances independently for each star, there exists spatial correlation in the prior on small scales. Consequently, if we have reason to believe that these stars belong to a particular stellar cluster, it's not advisable to rely solely on a combined distance estimation. Instead, it's more reliable to establish a model, treating the distance of the cluster as an adjustable parameter, and then solve it utilizing the parallaxes, taking into account their spatial correlations as well.

The posterior in Eq. \ref{eq:bj2} provides a detailed depiction of the distance to the source star. To quantify the level of uncertainty, \citet{2018AJ....156...58B} employed the highest density interval (HDI) with a specified probability $p$. The HDI represents the range of distances that encloses the region with the highest probability density in the posterior, and the integral of this region corresponds to the probability $p$. This interval is bounded by two values: $r_{lo}$ as the lower bound and $r_{hi}$ as the upper bound. In this context, \cite{2018AJ....156...58B} set the probability $p$ to be 0.6827, which is equivalent to the probability contained within ±$\sigma$ of the mode in a Gaussian distribution, $r_{mode}$, they refer to it as $r_{est}$. It is important to note that the distance posterior is often asymmetric, and the difference between $r_{hi}$ and $r_{est}$ as well as $r_{est}$ and $r_{lo}$ may not be equal, reflecting the asymmetric nature of the uncertainty in distance estimation, sometimes significantly so.

The HDI exhibits uniqueness when the posterior distribution is unimodal. In cases of multimodality, where the distribution may have multiple peaks, the HDI can split into several intervals \citep{b7f71c99-f621-3c7e-a9dd-9d152d4822a4}. When dealing with a specific parameter space (${\overline{\omega}}$, $\sigma_{\overline{\omega}}$, $L$), there are instances, approximately 0.09 percent of the time, where the posterior becomes bimodal \citep{2015PASP..127..994B}. In such scenarios, we preserve both the HDI and the mode estimator, ensuring uniqueness by excluding the minimum within the span. If the HDI cannot maintain uniqueness, we opt to utilize the median of the distribution as the point estimator. \citet{2018AJ....156...58B} reported that the $16^{th}$ and $84^{th}$ percentiles, forming the equal-tailed interval (ETI). This interval balances the probability above and below it, denoted as (1 $\pm$ $p$)/2.

Due to the absence of an analytic solution for the HDI in this posterior, \citet{2018AJ....156...58B} employed an iterative procedure involving a Taylor expansion. The process involves normalizing the posterior using Gaussian quadrature, denoted as $P(r)$. Starting at the mode (where the first derivative is zero) with a fixed negative $\Delta P$, \citet{2018AJ....156...58B}  compute an initial step size,

\begin{figure}
\centering
\begin{subfigure}[b]{0.46\textwidth}
\centering
\includegraphics[width=\textwidth]{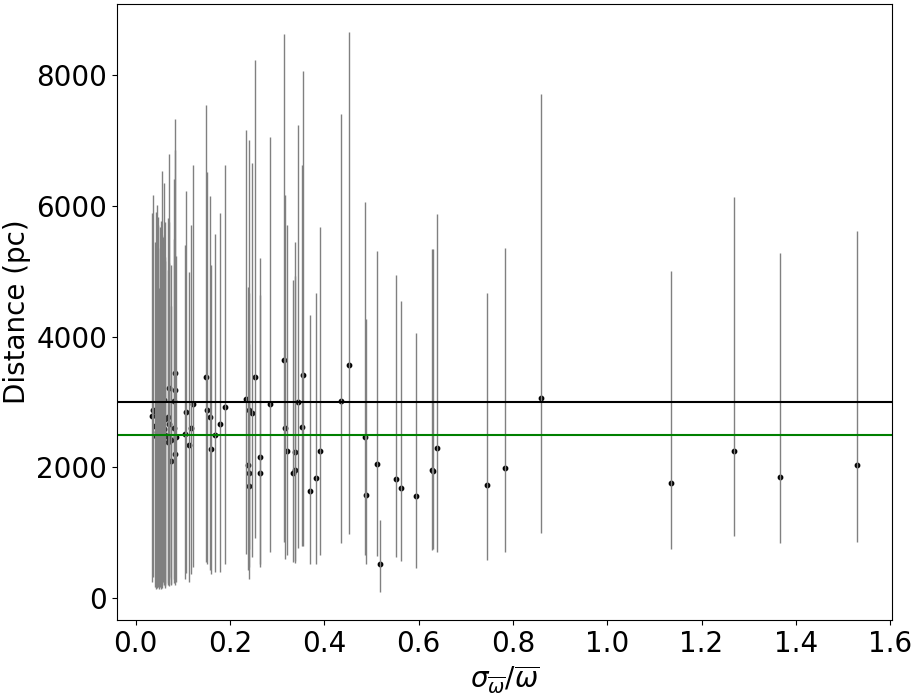}
\end{subfigure}
\caption{Plot of the
estimated distances, $r_{est}$, of the cluster members (black dots), against the $\sigma_{\overline{\omega}}$/${\overline{\omega}}$, with the error bars drawn as grey lines. The solid green line represents the mean distance of the IC 1590 cluster.}
\label{fig:dist}
\end{figure}

\begin{equation}
\Delta r_0 = \sqrt{2 \Delta P \left(\frac{d^2P}{dr^2}\right)_{r_{\text{mode}}}}
\label{eq:dis}
\end{equation}
The candidate's lower and upper bounds are defined as follows,
\begin{equation}
r^{\pm}_1 = r_{\text{mode}} \pm \Delta r_0
\label{eq:dis}
\end{equation}
A negative $\Delta P$ likely indicates that the algorithm starts by moving away from the mode in the direction of decreasing probability density. This could be a reasonable approach to exploring the tails of the distribution, which is essential for identifying the boundaries of the HDI. The iterative procedure continues by taking small steps away from the mode, computing the area under the curve covered by each step until the total area reaches the desired limit \citep{2018AJ....156...58B}.

When the posterior exhibits a bimodal distribution, which is determined separately from searching for roots where $dP/dr$ = 0, the process of finding the HDI starts from the peak with the highest value. By ensuring the search does not encounter the minimum between the two modes, the HDI remains well-defined \citep{2018AJ....156...58B}.
For the distance estimation, \citet{2018AJ....156...58B} utilized a fixed $\Delta P$ of - 0.01$P$($r_{\text{mode}}$) probably due to statistical or computational context, where $\Delta P$ is being adjusted based on the characteristics of the posterior distribution.
For 98 percent of the sources in $Gaia$ DR2, between 39 and 73 iterations
(the $1^{st}$ and $99^{th}$ percentiles) are required and the quantiles are found numerically by drawing 2×10$^{4}$ samples from the
posterior with a Markov Chain Monte Carlo (MCMC) method.\\
We applied this method for 91 identified cluster members and estimated the distance of each cluster member in Fig. \ref{fig:dist}. We plotted the estimated distances, denoted as $r_{est}$, of the cluster members as black dots, against the $\sigma_{\overline{\omega}}$/${\overline{\omega}}$. Each star is accompanied by error bars representing the confidence interval, with upper and lower bounds denoted as $r_{hi}$ and $r_{lo}$, respectively. The solid green line denotes the mean distance of the IC 1590 cluster and is $\overline r_{\text{est}}$ = $2503^{\text{+}711}_{-445}$ pc. The solid black line represents the reciprocal of the mean parallax for all 91 members of the IC 1590 cluster and is $\overline r_{\overline{\omega}}$ = 3012 $\pm$ 151 pc.

\section{Structure Parameters Estimation of IC 1590}
\label{sec:king}
The radial surface density profile of an open cluster is an important tool for examining the distribution of cluster members within the cluster and determining the extent or boundaries of the open cluster in the celestial coordinates ($\alpha-\delta$) plane, offering a clearer perspective on its size and shape \citep{10.1093/mnras/stac2116}. The profile can be fitted using a \cite{1962AJ.....67..471K} fit of the following form \citep{2019A&A...627A.119C,2022A&A...659A..59T},
\begin{equation}
\rho(r) = \rho_{0} \left(\frac{1}{\sqrt{1 + \left(\frac{r}{r_c}\right)^{2}}} - \frac{1}{\sqrt{1 + \left(\frac{r_t}{r_c}\right)^{2}}}\right)
\label{eq:king}
\end{equation}
where $\rho_0$ represents the core density and $\rho_{bg}$ represents the background density, respectively. The core radius of the cluster is represented by $r_c$; whereas $r_t$ represents the tidal radius, the value of $r$ at which $\rho = \rho_{bg}$. The radial density profile is derived by calculating the radial distance ($r$) of an $i^{th}$ cluster member ($\alpha_{i}$–$\delta_{i}$) from the center of the cluster  ($\alpha_{0}$–$\delta_{0}$) using,
\begin{equation}
\cos r = \cos \delta_{i} \cos \delta_{0} \cos (\alpha_{i}-\alpha_{0})+\sin \delta_{i} sin \delta_{0}\nonumber,
\label{eq:dis}
\end{equation}
where $i$ ranges from 1 to $N$. N represents the total number of members in the cluster. The stellar surface density, the number of stars per square arcmin ($\rho_{i}$) is derived as $\rho_{i}$ = $N_i$ /$A_i$. Here, $N_i$ represents the number of stars in the $i^{th}$ ring with inner and outer radii of $r_i$ and $r_{i+1}$, respectively, and $A_{i}$ = $\pi(r_{i+1}^2 - r_i^2)$ is the area of the ring. The density uncertainty for each ring is determined by $\sigma_{pi} = \sqrt{N_i}/A_i$, assuming Poisson statistics \citep{10.1093/mnras/stac2116}.

\begin{figure}
\centering
\begin{subfigure}[b]{0.45\textwidth}
\centering
\includegraphics[width=\textwidth]{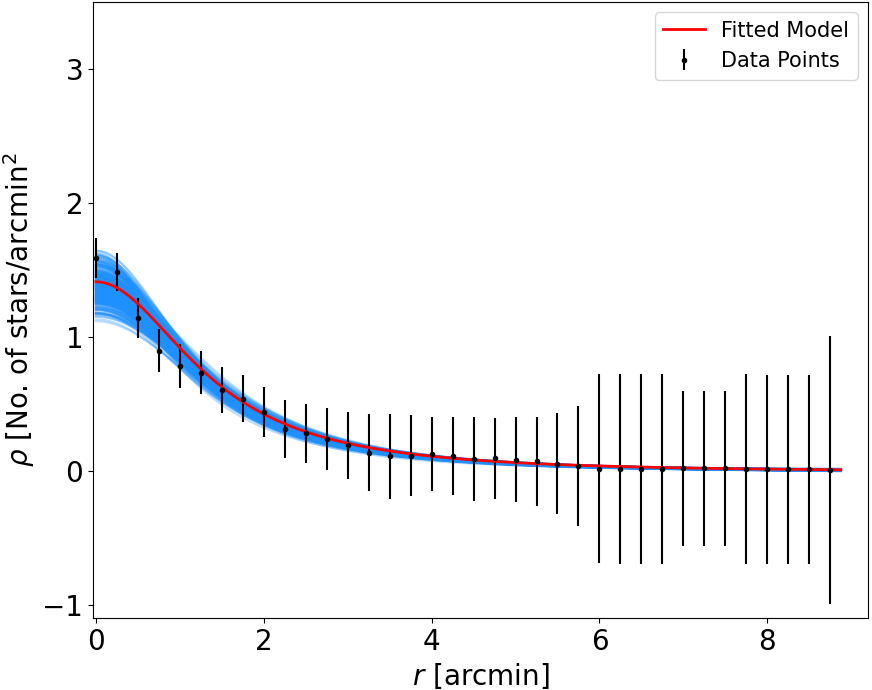}
\end{subfigure}
\caption{Plot of results of the radial density profile fit for IC 1590 for cluster member within a radius of 9 arcmin. The error bars indicate the range of uncertainty, which is $\pm1$$\sigma$ Poissonian. The MCMC chains generated 300 random sample fits, represented by blue lines, to illustrate model parameter uncertainties.}
\label{fig:rad}
\end{figure}
\begin{figure}
\centering
\begin{subfigure}[b]{0.47\textwidth}
\centering
\includegraphics[width=\textwidth]{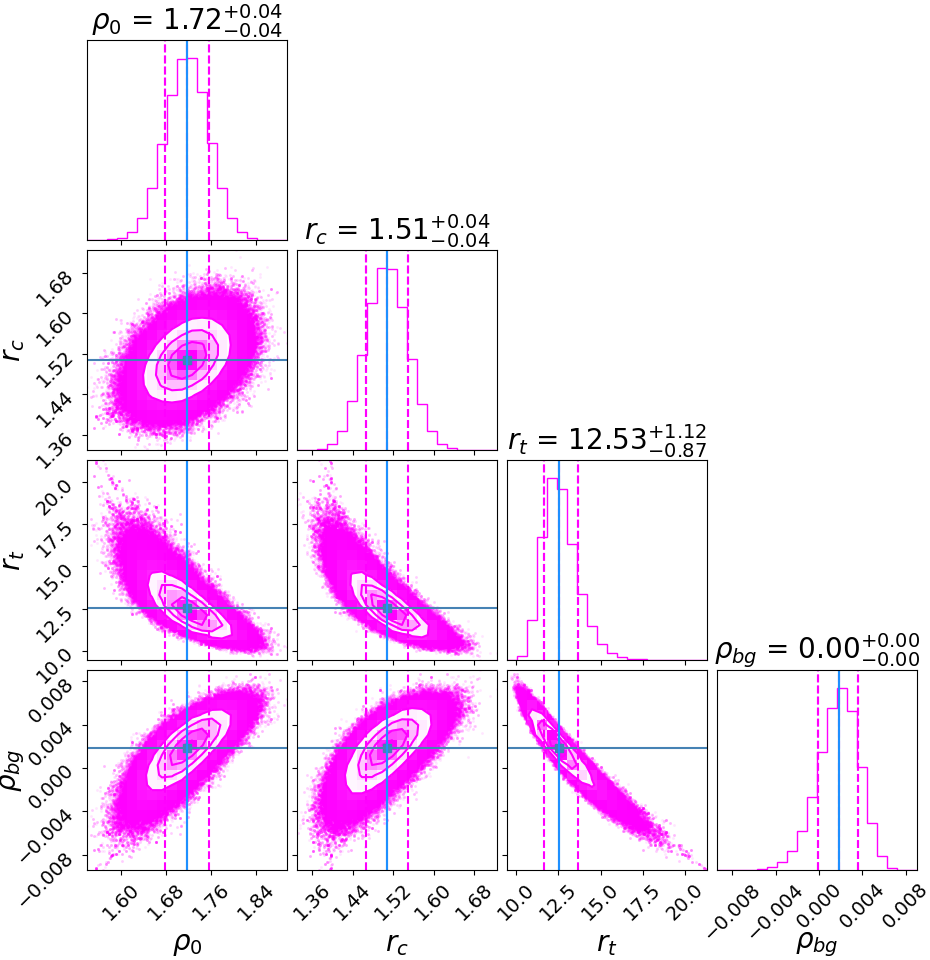}
\end{subfigure}
\caption{MCMC analysis used to obtain the marginalized posterior distribution and uncertainties associated with the model parameters of the King profile fit for IC 1590 for cluster member within a radius 9 arcmin. The statistical uncertainties indicate the differences between the mean value and the $16^{th}$ percentile (lower limit) and $84^{th}$ percentile (upper limit). The magenta dashed lines represent the $16^{th}$, $50^{th}$, and $84^{th}$ percentiles.}
\label{fig:mc}
\end{figure}
The radial density profile is fitted using Eq. \ref{eq:king} for 91 cluster members of IC 1590. In an ideal scenario, the number density of background stars, $n_{bg}$ should be zero, implying no contamination from the background in the membership analysis. However, to quantify and determine the degree of contamination for the cluster, a small non-zero value of $\rho_{bg}$ is utilized \citep{2022A&A...659A..59T}. The Bayesian Markov Chain Monte Carlo (MCMC) method is used to determine the values and the uncertainties of the parameters. For MCMC sampling, 2000 walkers, 1000 iterations, and 200 burn-in steps are utilized \citep{10.1093/mnras/stac2116}. There is no fixed rule for setting these numbers, and it often involves a trial-and-error process based on the characteristics of the specific problem and the behavior of the MCMC algorithm. The marginalized posterior distributions for each parameter are shown in Fig. \ref{fig:mc}. The distribution's high symmetry indicates that the mean and median are almost the same. The mean value, which is the $50^{th}$ percentile of the distribution, is used as the best-fitting parameter value. The statistical uncertainties are calculated by determining the values both up and down to the $16^{th}$ and $84^{th}$ percentiles of the distribution. textcolor{teal}{In Fig. \ref{fig:rad}, the observed data is represented by black dots. The line that best fits the data is shown in red. Additionally, 300 random sample fits generated from the MCMC chains are represented by the blue lines in Fig. \ref{fig:rad}, providing a visual representation of the limit of uncertainties associated with the model parameters.
The core radius is found to be $r_c$ = 1.51 ± 0.04 arcmin, tidal radius $r_t$ = $12.53^{\text{+}1.12}_{-0.87}$ and central density $\rho_0$ = 1.72 ± 0.04 stars/arcmin$^2$.

\section{Orbit Analysis of IC 1590}
\label{sec:orb}
The open clusters are known to constitute one of the best classes of objects to investigate the Galactic structure and Stellar Dynamics \citep{10.3389/fspas.2021.656474}. Understanding the kinematics and dynamics of open clusters is crucial in determining their birth radii and their distribution across the galaxy. This information can provide insights into the formation and evolution of these objects as well as their role in the larger picture of star formation in our galaxy. The determination of their fundamental parameters as radial velocities, distances, and ages allows us to present the analyses of the distribution of the open clusters in the solar neighborhood, exploring the birthplaces and actual positions on the Galactic plane \citep{2023AJ....165...79Y}. We used MWPotential2014 in the Galactic dynamics library \href{https://github.com/jobovy/galpy}{galpy} python package \citep{2015ApJS..216...29B} for orbit analysis of IC 1590, which assumes an axisymmetric potential for the Milky Way. The MWPotential2014 model represents the composite of three distinct components in the Galactic potential. These components include the spherical bulge as described in Bovy \cite{2015ApJS..216...29B}, the Galactic disc as defined by \cite{1975PASJ...27..533M}, and the massive, spherical dark-matter halo as defined by \cite{1996ApJ...462..563N}. In our analysis, we adopted the Galactocentric distance and orbital velocity of the Sun as $R_{gc}$ = 8 Kpc and $V_{rot}$ = 220 km s$^{-1}$, respectively \citep{2015ApJS..216...29B,2012ApJ...756...89B}. The distance of the Sun from the Galactic plane was accepted as 25 ± 5 pc \citep{2008ApJ...673..864J}.

To perform a complete orbit integration, it is necessary to know the radial velocity parameter. We have adopted the mean radial velocity of IC 1590 as $V_{\gamma}$ = -32.46 ± 6.36 km s$^{-1}$ from \cite{2007AN....328..889K}. We provided the following parameters as input to perform the orbit integration: the central equatorial coordinates ($\alpha$ = 13.205, $\delta$ = 56.629) in degrees adopted from \citet{2020A&A...633A..99C}, the mean proper-motion components, $\mu_{\alpha}$ cos $\delta$ = -2.37 ± 0.07 and $\mu_{\delta}$ = -1.49 ± 0.08 mas yr$^{-1}$ from Table \ref{tab:gmm}, the isochrone distance, $d_{iso}$ = 2.87 ± 0.02 Kpc from Section \ref{sec:isodis}, and the radial velocity, $V_{\gamma}$ = -32.46 ± 6.36 km s$^{-1}$ from \cite{2007AN....328..889K}. We performed a forward integration of the cluster's orbit, using a step size of 1 Myr, covering a period of up to 6 Gyr to determine the potential current location of the cluster. The "side view" of the cluster on the $Z\times R_{gc}$  plane is shown in Fig. \ref{fig:orb}, providing information about its distance from both the Galactic plane and the Galactic center. Here, $Z$ is the vertical distance from the galactic plane, and $R_{gc}$ is the Galactocentric distance of the Milky Way. To estimate the possible birth radius of IC 1590, we carried out orbit analyses for a past epoch equivalent to its age, which is 3.54 Myr. The integration procedure was constrained to the age of the cluster due to the presence of potential-based uncertainties in time and additional errors associated with factors like distance, proper motion components, and radial velocity \citep{2021AJ....161..101S}. Fig. \ref{fig:birth} shows the distance of the cluster on the $R_{gc}\times t$ plane with time, where $t$ is the age of the cluster.

Orbit integration yielded the following parameters for IC 1590: birth radius ($R$ = 9.853 ± 0.059 Kpc), apogalactic distance($R_a$ = 9.858 ± 0.394 Kpc), and perigalactic distance
($R_p$ = 9.856 ± 0.054 Kpc), eccentricity ($e$ = 0.00086 ± 0.0099), maximum vertical distance from the Galactic
plane ($Z_{max}$ = 317 ± 54 pc) and orbital period ($P_{orb}$ = 447 ± 5 Myr). Perigalactic and apogalactic distances emphasize that the orbit of IC 1590 is completely outside the solar circle as shown in Fig. \ref{fig:orb}, which shows that IC 1590 was born
in a metal-poor region \citep{2023PARep...1....1T,2023AJ....165...79Y}. Hence, the cluster reaches a maximum vertical distance $Z_{max}$ =  317 ± 54 pc above the Galactic plane, suggesting that IC 1590 belongs to the thin-disc component of the Milky Way.
\begin{figure}
\centering
\begin{subfigure}[b]{0.46\textwidth}
\centering
\includegraphics[width=\textwidth]{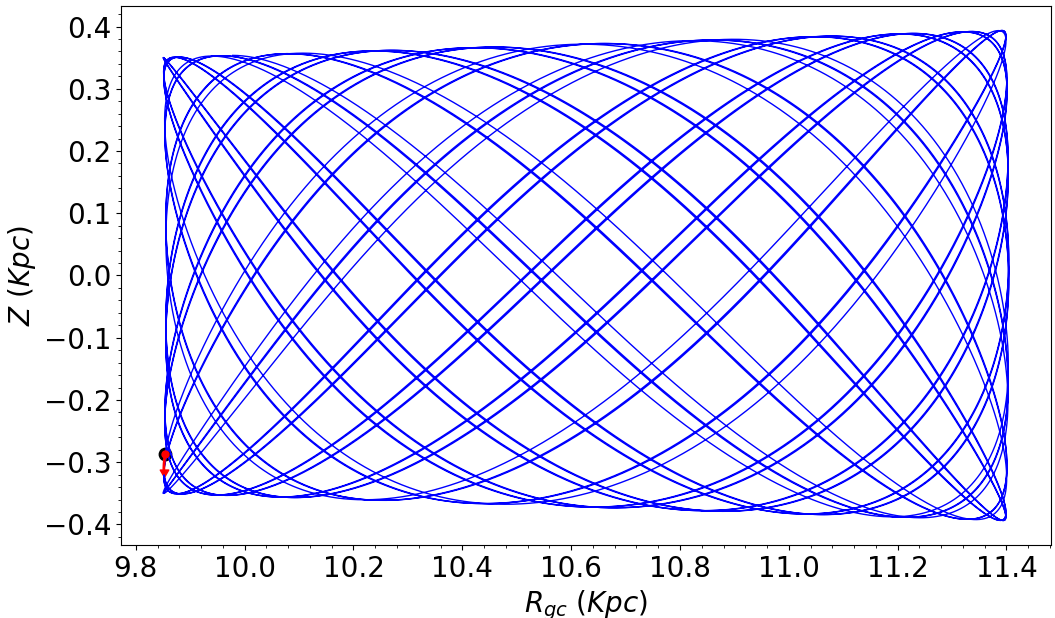}
\end{subfigure}
\caption{The Galactic orbits and birth radii of IC 1590 in the $Z$$\times$$R_{gc}$ planes. The red circle represents the
present-day position of IC 1590. The red arrow is the motion vector of IC 1590.}
\label{fig:orb}
\end{figure}
\begin{figure}
\centering
\begin{subfigure}[b]{0.46\textwidth}
\centering
\includegraphics[width=\textwidth]{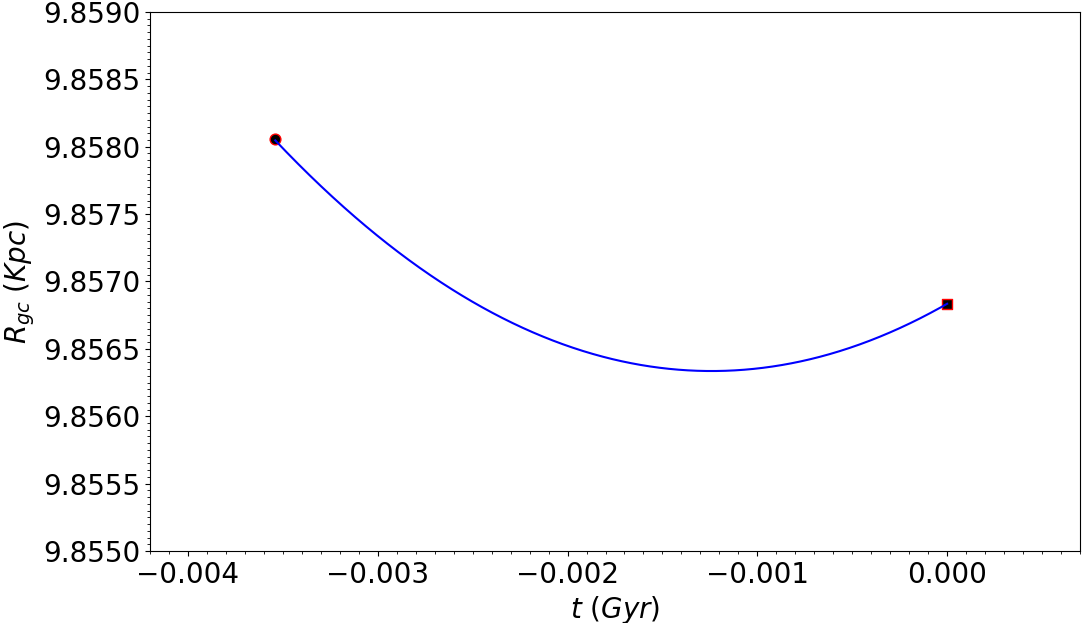}
\end{subfigure}
\caption{The Galactic orbits and birth radii of IC 1590 in the $R_{gc}$$\times$$t$ planes. The black circle represents the birth locations and the black square represents the present-day position of IC 1590.}
\label{fig:birth}
\end{figure}
\section{Results}
\label{sec:res}
Applying an unsupervised ensemble machine learning approach \citep{10.1093/mnras/stac2116} to IC 1590 using $Gaia$ DR3 astrometric data, we identified 91 stars as cluster members within a 9 arcmin search radius for stars with membership probability value of > 0.50. We used Gaussian fits to analyze the distributions of parallax and proper motions ($\pi$, $\mu_{\alpha}$$\cos \delta$, $\mu_{\delta}$), to obtain the mean values of the parallax, proper motion in right ascension, and proper motion in declination as $\pi$ = 0.327 ± 0.07 mas, $\mu_{\alpha}$$cos$ $\delta$ = -2.37 ± 0.12 mas yr$^{-1}$ and $\mu_{\delta}$ =  -1.49 ± 0.09 mas yr$^{-1}$, respectively. The directions of proper motions ($\mu_{\alpha}$$cos$ $\delta$, $\mu_{\delta}$) of the 91 cluster stars at their positions ($\alpha$, $\delta$) as shown in Fig \ref{fig:gmm6}. We can see that almost all the cluster members exhibit motion in the same direction, indicating a very effective determination of their membership in the cluster.

We estimated the best-fitted parameters for IC 1590 from isochrone fitting using \href{https://github.com/asteca}{ASteCA} based on $Gaia$ EDR3 to be distance $d$ $\sim$ 2.87 ± 0.02 Kpc, age $\sim$ 3.54 ± 0.05 Myr, metallicity $z$ $\sim$ 0.0212 ± 0.003, binarity value of $\sim$ 0.558 and extinction $A_v$ $\sim$ 1.252 ± 0.4 mag for an $R_v$ value of $\sim$ 3.322 ± 0.23. This result is in good agreement with the distance calculated from the $Gaia$ DR3 parallax $d$ = 3.08 ± 0.15 Kpc and also consistent with the distance $d$ = 2.82 ± 0.24 kpc from radio VLBA astrometry of an H$_2$O maser in NGC 281 West \citep{2008PASJ...60..975S}. The position of the PMS on the isochrone provides an estimate of their age. So, we estimate the age of the cluster from PMS isochrone to be 3.54 ± 0.29 Myr which is well consistent with the prediction from the theoretical evolutionary model for contracting PMS given by \citet{1993ApJ...418..414P}.\\
We estimated the slope of the $G$-band LF slope = 0.33 ± 0.09 and slope of the MF $\alpha$ = 1.081 ± 0.112 for a total mass of $M_T$ = 255 M$\sun$ when binary stars are considered as single stars and $\alpha$ = 1.490 ± 0.051 for a total mass of $M_T$ = 152 M$\sun$ taking into account the binarity value of $\sim$ 0.558 estimated by \href{https://github.com/asteca}{ASteCA}. \citet{1997AJ....113.2116G} have reported the slope of the MF ( 1.00 ± 0.21) and  Sharma et al. (2012) reported (1.08 ± 0.15 to 1.16  ± 0.15) for the NGC 281 region which, within error, is comparable to the value that obtained in our study. \citet{2021AJ....162..140K} also reported that the slope of the IMF of the synthetic cluster for IC 1590 taking into account the resolved binary stars is $\sim$ 1.30 ± 0.04. We estimated the distance of individual stars from $Gaia$ DR2 parallaxes using \citet{2018AJ....156...58B} method and the mean distance was $\sim$ $2503^{\text{+}711}_{-445}$ pc.

From the 2MASS $JHK_s$ data, we found 68 counterparts for cluster members shown by open circles in Fig. \ref{fig:tcd}. From the 3rd MSFRs Omnibus X-ray Catalog (MOXC3), we found 42 counterpart X-ray sources for cluster members, and from the SOS. VII. UBVI Photometry Data, we found 4 H$_{\alpha}$ sources as shown in Fig. \ref{fig:tcd}. The $JHK_s$ colors were transformed from the 2MASS system to the Koornneef system using the relations given by \citet{2001AJ....121.2851C}. Using $(J-H)$ versus $(H-K)$ color-color diagram we identified the O-type stars, X-ray sources, and NIR-excess sources in the cluster region as shown in Fig. \ref{fig:tcd} as different types of stellar objects occupying distinct regions on the CCs.\\
From the NIR $J$ versus $(J-H)$ CMD in Fig. \ref{fig:ccm}, we found that the identified NIR-excess YSOs are probably the pre-main sequence (PMS) stars of ages $\leq$ 2 Myr and masses $\sim$ 0.35 - 5.5 M\sun.  Most of the PMS stars are undergoing evolution along Hayashi tracks or in the Kelvin–Helmholtz contraction phase \citep{2021AJ....162..140K}. From NIR CMD, it is found that the identified YSOs are probably PMS stars of ages $\leq$ 1 Myr and the majority of these stars have masses between 0.5 - 3.5 M$\sun$ \citep{2012PASJ...64..107S}. It is also reported that the upper age limit of these stars is 1.9 Myr and the upper mass limit of these PMS stars appears to be 5 M$\sun$ \citep{2021AJ....162..140K}.

We have fitted the radial surface density profile using \cite{1962AJ.....67..471K} fit based on the MCMC method and we estimated the core radius $r_c$ = 1.51 ± 0.04 arcmin, tidal radius $r_t$ = $12.53^{\text{+}1.12}_{-0.87}$ arcmin and central density $\rho_0$ = 1.72 ± 0.04 stars/arcmin$^2$.  The core radius is well consistent with the value obtained by \citet{2012PASJ...64..107S} which is $r_c$ = 1.7 ± 0.4 arcmin.\\
The analysis of galactic orbits has revealed that IC 1590 follows a distinctive boxy trajectory outside the solar circle, indicating that it is associated with the young thin-disc component of the Milky Way. Moreover, the birth radius (9.853 ± 0.059 Kpc) suggests that this cluster originated beyond the solar circle within the Milky Way.\\
We have also generated extinction maps using the  \href{http://smeingast.github.io/PNICER/ }{PNICER} technique. The extinction in this region is estimated to be $A_v$ = 0 - 1.45  mag. The extinction map is analogous to the density structure revealed by the contours extracted from the 500 $\mu$$m$ dust continuum emissions of \textit{Herschel} SPIRE.
\section{Summary and Conclusion}
\label{sec:sum}
\begin{figure}
\centering
\begin{subfigure}[b]{0.45\textwidth}
\centering
\includegraphics[width=\textwidth]{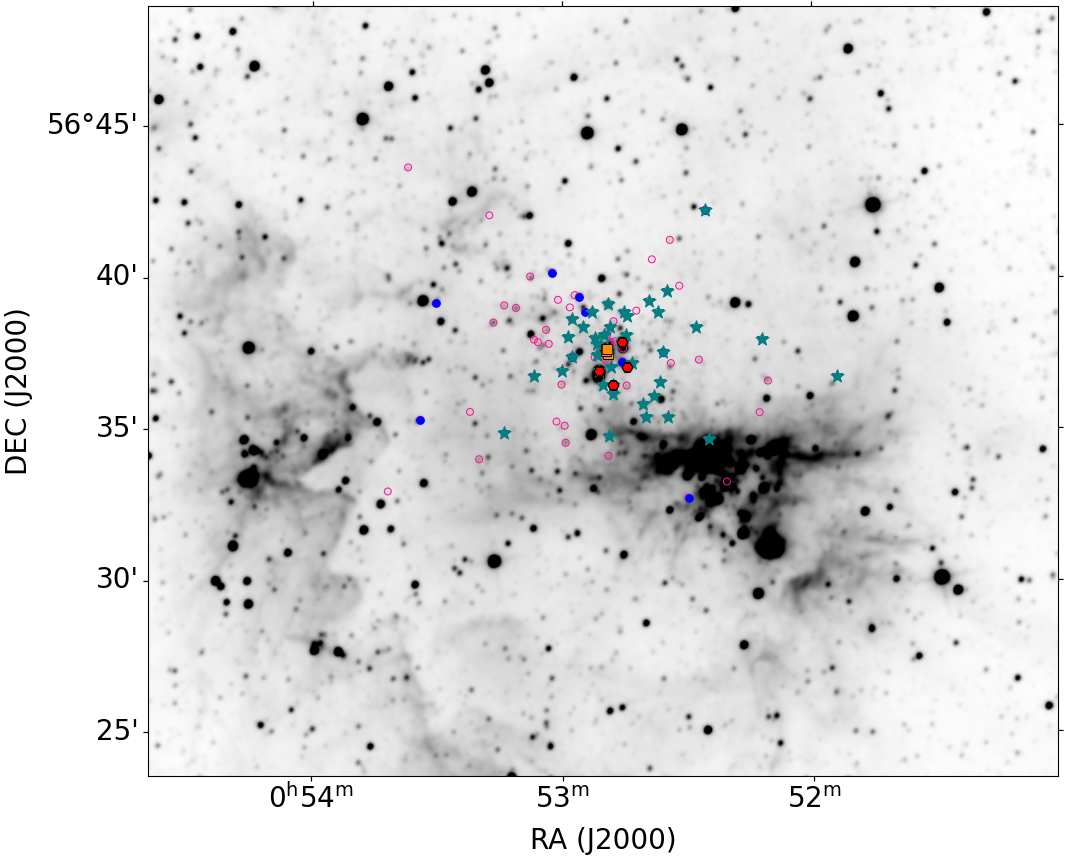}
\label{fig:1}
\end{subfigure}
\caption{The 4.63 {$\mu$}$m$ MIR image of NGC 281 taken from the all-WISE. We represent all the possible cluster members (pink open circles), X-ray counterparts (teal star symbols), O-type stars (orange-filled square symbols), NIR-excess YSOs (blue-filled circles), and H$_{\alpha}$ sources  (red-filled hexagon symbols) in the cluster region.}
\label{fig:w4}
\end{figure}
\begin{figure}
\centering
\begin{subfigure}[b]{0.45\textwidth}
\centering
\includegraphics[width=\textwidth]{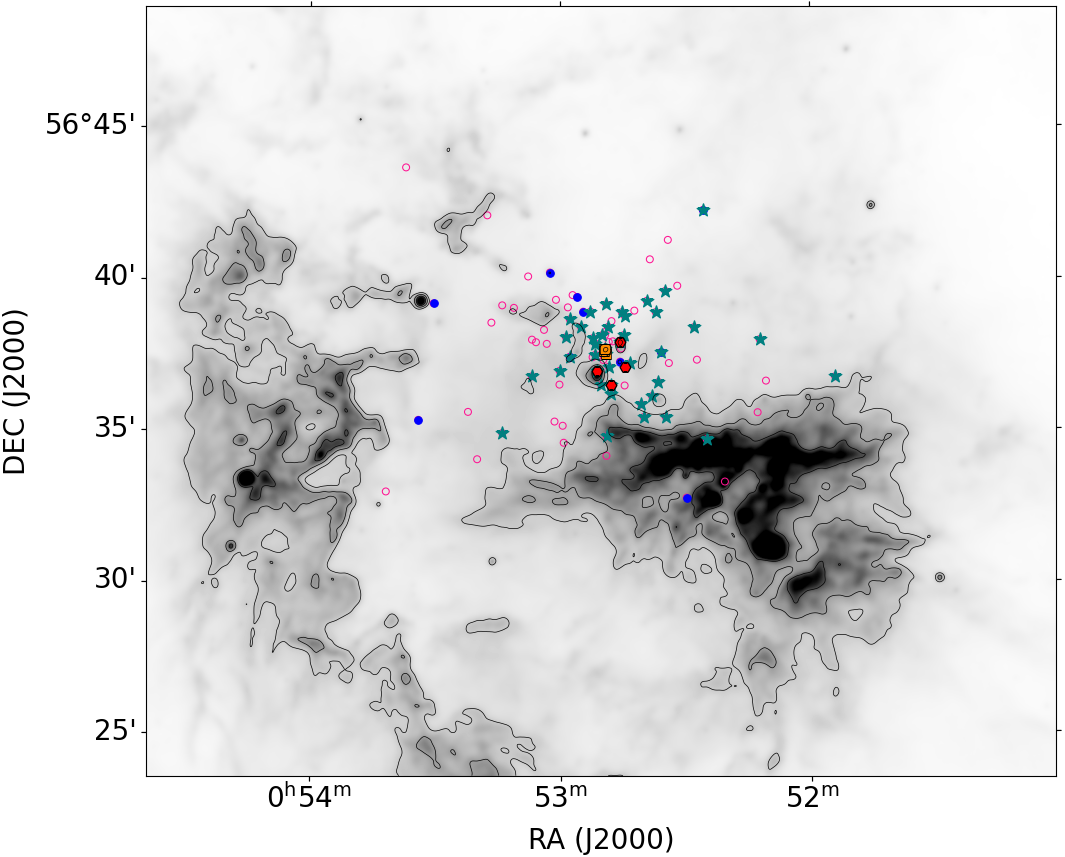}
\end{subfigure}
\caption{The 12 {$\mu$}$m$ MIR image of NGC 281 taken from the all-WISE. We represent all the possible cluster members (pink open circles), X-ray counterparts  (teal star symbols), O-type stars (orange-filled square symbols), NIR-excess YSOs (blue-filled circles), and H$_{\alpha}$ sources  (red-filled hexagon symbols) in the cluster region. The contours are extracted from all-WISE 12  {$\mu$}$m$ dust continuum emissions.}
\label{fig:w12}
\end{figure}
Young open clusters are found to contain some emission stars that influence the star formation in that star-forming region. The 4.63 and 12 {$\mu$}$m$ mid-infrared (MIR) images were taken from the Wide-field Infrared Survey Explorer (all-WISE) \citep{2010AJ....140.1868W} as shown in Fig. \ref{fig:w4} and Fig. \ref{fig:w12}, respectively. It shows that there are some irregular structures present near the cluster region. The distribution of young stellar objects (YSOs) identified based on their NIR excess suggests that star formation in and around the cluster is not happening at the same time or in a synchronized manner. The distribution of these structures, along with the ages and NIR-excess YSOs and the T Tauri Stars from color-magnitude diagrams, indicates that star formation has been influenced or triggered by the presence of the cluster in this star-forming region. In the cluster region, pre-main sequence stars are evenly distributed throughout, while the Be stars are observed to be concentrated nearer to the central region of the cluster. This observation suggests a potential trend where more massive stars tend to preferentially form closer to the center of the cluster \citep{2006MNRAS.370..743S}.\\
\cite{2005ASPC..341..107H} have suggested the following evidence
indicating star formation in a region may have occurred due
to an external trigger: (i) The presence of a significant number of YSOs concentrated in the vicinity of compressed gas within HII regions. (ii) These YSOs tend to exhibit a range of ages, with some extending up to the age of the massive ionizing stars within the region. If star formation occurred independently of the influence of massive stars, one would expect an absence of correlation between star formation and the compressed gas around HII regions. In such cases, regions rich in young stars should appear in extended areas beyond the reach of the expanding HII regions. Furthermore, the ages of YSOs would typically exceed the ages of the massive ionizing stars \citep{2007MNRAS.379.1237M}. The sources are closer to HD 5005 with ages of
$\leq$ 2 Myr which is less than the age $\sim$ 3.54 Myr that is estimated for IC 1590. This is consistent with the scenario given by \cite{2005ASPC..341..107H} and represents a strong endorsement for the models of triggered star formation towards IC 1590.

The NGC 281 region contains a Trapezium-like system HD 5005, which consists of four O-type stars at the center \citep{2011ApJS..193...24S}. There are two molecular clumps associated with ionized hydrogen: an eastern clump (NGC 281 East) and a western clump (NGC 281 West). NGC 281 West is more massive and compact compared to the elongated NGC 281 East \citep{2012PASJ...64..107S}. The main ionizing source HD 5005 is located to the northeast of NGC 281 West and northwest of NGC 281 East \citep{2012PASJ...64..107S}. \citet{2003NewA....8..191L} mapped the NGC 281 West both in \ce{^{12}C}O (J = 1 - 0)  and \ce{^{13}C}O (J = 1 - 0) emission lines and found a close association between
the \ce{^{12}C}O (J = 1 - 0) emission peak and the H$_2$O maser source, indicating the ongoing star formation \citep{1978ApJ...219..467E}. The differential extinction towards the central cluster is approximately 0.2 mag \citep{2012PASJ...64..107S}, indicating that the central cluster region contains only gas and dust of low density, likely due to the effects of the massive stars at the cluster center. The western molecular clump of NGC 281 is interacting with the ionized gas \citep{2003NewA....8..191L}. Based on kinetic evidence, \cite{1979ApJ...227L..93E} suggested the passage of a shock through NGC 281 West. There are also hints of an outflow near the maser source, suggesting active star formation in NGC 281 West \citep{1990ApJ...352..139S,1994ApJS...94..615H,1996A&AS..115..283W}. They further suggested that this area is a site of triggered star formation through the process known as "collect and collapse" \citep{2012PASJ...64..107S}. As a result, NGC 281 West has been proposed as a star-forming region triggered by IC 1590 \citep{2021AJ....162..140K}.

NGC 281 East, another star-forming region, was found through \ce{^{12}C}O  (J = 1 - 0) observations by \cite{1978ApJ...219..467E}. This region contains 3 IRAS sources, and a highly reddened star was found in the clump at the northern edge of NGC 281 East. \cite{1997AJ....114.1106M} suggested that the star formation in NGC 281 East may have been initiated or enhanced by shocks. The ionization and shock fronts created by high-mass stars from the first generation have sparked the formation of a new generation of stars at the boundaries of the molecular clumps \citep{2012PASJ...64..107S}. \citet{2002ASPC..267..257M,2003RMxAC..15..151M} also suggested that the NGC 281 complex was formed in a fragmenting superbubble. Nonetheless, there is still a need to investigate the physical connections between these star-forming regions and IC 1590.

\section*{Acknowledgements}
The authors thank Prof. Michael Stanley Bessell for his valuable comments and suggestions that helped us to improve the quality of our paper. We made use
of NASA’s Astrophysics Data System as well as the VizieR and
Simbad databases at CDS, Strasbourg, France, and data from the
European Space Agency (ESA) mission $Gaia$, processed by
the $Gaia$ Data Processing and Analysis Consortium (DPAC). We highly acknowledge the $Herschel$ Science Archive from which we have downloaded
$Herschel$ SPIRE 500 $\mu$$m$ dust emission map for NGC 281 region. We also make use
of data from the 2MASS, which is
a joint project of the University of Massachusetts and the Infrared
Processing and Analysis Center/California Institute of Technology,
funded by the National Aeronautics and Space Administration and the National Science Foundation. This work has made use of $Aladin$ $sky$ $atlas$ developed at CDS, Strasbourg Observatory, France. We also make use of data
products from the all-WISE, which is a
joint project of the University of California, Los Angeles, and the Jet
Propulsion Laboratory/California Institute of Technology, funded by
the National Aeronautics and Space Administration. We also used TOPCAT, an interactive graphical viewer and editor for tabular data.

\section*{Data Availability}
The data used in this paper will be made available upon reasonable request to the corresponding author.



\bibliographystyle{mnras}
\bibliography{example} 





\bsp	
\label{lastpage}
\end{document}